\documentclass[11pt,a4paper]{article}
\pdfoutput=1

\usepackage{jheppub}
\usepackage{hyperref}
\usepackage{booktabs}
\usepackage{xspace}
\usepackage{xcolor}
\usepackage[toc,page]{appendix}
\usepackage{mathtools}
\usepackage[T1]{fontenc}

\usepackage[normalem]{ulem}

\newcommand{\noun}[1]{{\tt #1}}
\newcommand{\MSbar}{\overline{\rm MS}}
\newcommand{\POWHEG}{\noun{POWHEG}}

\newcommand{\POWHEGBOX}{\noun{POWHEG-BOX}}

\newcommand{\POWHEGBOXVTWO}{\noun{POWHEG-BOX-V2}}

\newcommand{\MINLO}{\noun{MiNLO}}

\newcommand{\hbbgMINLO}{\noun{Hbbg-MiNLO}}

\newcommand{\MCFM}{\texttt{MCFM-8.0}}

\newcommand{\PYTHIA}[1]{\noun{Pythia{#1}}}

\newcommand{\HZ}{\textrm{HZ}}

\newcommand{\as}{\alpha_{\scriptscriptstyle s}}

\newcommand{\murdec}{\mu_{\scriptscriptstyle \mathrm{R,dec}}}
\newcommand{\mur}{\mu_{\scriptscriptstyle \mathrm{R}}}

\newcommand{\bb}{b\bar{b}}
\newcommand{\sdec}{M_H^2}

\newcommand{\gev}{{\rm{GeV}}}

\newcommand{\hbb}[1]{\ensuremath{{\rm{H}}\to b\bar{b}{#1}}}
\newcommand{\yres}{\ensuremath{y_3}}
\newcommand{\ycut}{\ensuremath{y_{\rm{cut}}}}
\newcommand{\Gam}{\ensuremath{ \Gamma_{\hbb{}} }}
\newcommand{\GamMINLO}{\ensuremath{ \Gamma_{\hbb{}}^{\MINLO} }}
\newcommand{\GamMINLOa}{\ensuremath{ \Gamma_{\hbb{}}^{\MINLO,\rm{A}} }}
\newcommand{\GamMINLOb}{\ensuremath{ \Gamma_{\hbb{}}^{\MINLO,\rm{B}} }}
\newcommand{\GamLO}{\ensuremath{ \Gamma_{\hbb{}}^{\rm{LO}} }}
\newcommand{\GamNLO}{\ensuremath{ \Gamma_{\hbb{}}^{\rm{NLO}} }}
\newcommand{\GamNNLO}{\ensuremath{ \Gamma_{\hbb{}}^{\rm{NNLO}} }}
\newcommand{\Phibb}{\ensuremath{ \Phi_{b\bar{b}} }}
\newcommand{\Phibbg}{\ensuremath{ \Phi_{b\bar{b}g} }}
\newcommand{\evHfull}{{\tt{event\_Hprod\_x\_dec}}}
\newcommand{\evHprod}{{\tt{event\_Hprod}}}
\newcommand{\evHdec}{{\tt{event\_Hdec}}}
\newcommand{\scprod}{\tt{scalup\_prod}}
\newcommand{\scdec}{\tt{scalup\_dec}}
\newcommand{\ph}{p_{\rm{H}}}
\newcommand{\mh}{M_{\rm{H}}}
\newcommand{\mz}{M_{\rm{Z}}}
\newcommand{\mw}{M_{\rm{W}}}
\newcommand{\mbMSbar}{\overline{m}_b}
\newcommand{\BrHbb}{ {\rm{Br}}({\rm{H}}\!\rightarrow\! b\bar{b}) }
\newcommand{\mll}{M_{\ell\bar\ell}}
\newcommand{\mbb}{M_{b\bar b}}
\newcommand{\mbPY}{m_{b}^{\texttt{\scriptsize PY}}}

\newcommand{\omx}{1-x}

\newcommand{\lp}{\left(}
\newcommand{\rp}{\right)}
\newcommand{\lb}{\left[}
\newcommand{\rb}{\right]}

\preprint{ \\\\ TTP19-049 \\ LAPTH-052/19\\  MPP-2019-263 }

\title{{NNLOPS description of the H$\to\! b\bar{b}$ decay with \MINLO{}}}

\author[a,b]{Wojciech Bizo\'n,}
\author[c]{Emanuele Re,}
\author[d]{Giulia Zanderighi}

\affiliation[a]{Institut f{\"u}r Theoretische Teilchenphysik (TTP), KIT, 76128 Karlsruhe, Germany}
\affiliation[b]{Institut f{\"u}r Kernphysik (IKP), KIT, 76344 Eggenstein-Leopoldshafen, Germany}
\affiliation[c]{LAPTh, Universit\'e Grenoble Alpes, Universit\'e Savoie Mont Blanc, CNRS, 74940 Annecy, France}
\affiliation[d]{Max-Planck-Institut f\"ur Physik, F\"ohringr Ring 6, D-80805 Munich, Germany}

\emailAdd{wojciech.bizon@kit.edu}
\emailAdd{emanuele.re@lapth.cnrs.fr}
\emailAdd{zanderi@mpp.mpg.de}

\abstract{
  We present an event generator that describes the Higgs boson decay
  into $b$-quarks at next-to-next-to-leading-order (NNLO) in QCD and
  allows for a consistent matching to parton shower.
  For this purpose, we work within the \POWHEG{} framework and employ the
  \MINLO{} method to produce an NNLO-accurate \hbb{} event sample
  which can further be interfaced with a generic Higgs boson
  production mode.
  The resulting events, that combine Higgs production and decay, can
  be matched to parton shower by means of a vetoed shower.
  As an example, we present a short phenomenological study for
  associated Higgs production (HZ) and vector boson fusion (VBF)
  processes.
  We release the source code that was used to obtain these results as
  an {\texttt{h\_bbg}} user process in \POWHEGBOXVTWO{}. We also
  provide tools that enable interfacing the decay events with events for an
  arbitrary Higgs production mode and subsequent matching of the
  combined Les Houches (LH) events with a parton shower.
}


\begin{document}
\maketitle
\flushbottom

\section{Introduction}
\label{sec:intro}

Since the discovery of the Higgs boson by ATLAS and CMS in 2012, the
study of Higgs properties became a central part of the LHC physics
programme.
The mass of the Higgs boson is already measured with an accuracy of
about 0.2\%~\cite{Aad:2015zhl}, hence, within the Standard model (SM),
all couplings of the Higgs to other particles are predicted.
Nevertheless, beyond the Standard Model (BSM) effects can modify these
couplings, therefore constraining Higgs properties becomes one of the
priorities of the LHC physics programme.
For this reason, increasing experimental and theoretical precision of
measurements and predictions for production and decay rates is crucial
to enhance the sensitivity of searches for New Physics.

The largest branching fraction of the Higgs boson is to $b$-quarks.
Studies of very rare Higgs production modes obviously benefits from
the large \hbb{} branching fraction the most. For instance, in double-Higgs
production, which is the crucial process to constrain the Higgs
self-coupling directly, the most promising decay modes ($4b$,
$2b+2\tau$, $2b+2\gamma$) involve at least one Higgs boson decaying
into $b$-quarks. This is true both for the High-Luminosity (HL) and
High-Energy (HE) LHC~\cite{Cepeda:2019klc} programmes.

The \hbb{} decay mode is notably hard to study experimentally because
$b$-quarks typically give rise to $b$-jets which are more complicated
final states compared to decay modes involving charged
leptons. Furthermore, extracting the \hbb{} signal is challenging
because of large QCD backgrounds, coming from gluons splitting to
$b$-quarks and/or from other SM processes.  The sensitivity can be
enhanced by using the resonant kinematics of the signal, by requiring
$b$-jets with high transverse momentum and by exploiting different properties of jets
originating from a boosted colour singlet or a boosted gluon.
A lot of progress was achieved in this direction in recent years and
refined techniques were suggested and are currently used in
experimental measurements, see
e.g. Refs.~\cite{Altheimer:2013yza,Marzani:2019hun}.
Both ATLAS and CMS reported evidence for the \hbb{}
decay~\cite{Aaboud:2017xsd,Sirunyan:2017elk} followed by an
observation~\cite{Aaboud:2018zhk,Sirunyan:2018kst}.
Significant increase in statistics of the upcoming Run III and the
HL-LHC programme will allow for a comprehensive use of boosted
techniques and considerably improve the significance of these
measurements.

On the theoretical side, the computation of higher-order corrections
to the Higgs decay is quite advanced. The NLO QCD corrections to the Higgs decay to massive b-quarks have been
known for a long
time~\cite{Braaten:1980yq,Sakai:1980fa,Janot:1989jf,Drees:1990dq,Kataev:1992fe,Kataev:1993be}. NNLO corrections for massless $b$-quarks were computed more
recently~\cite{Anastasiou:2011qx,DelDuca:2015zqa} with even first
N$^3$LO results
appearing~\cite{Mondini:2019gid,Mondini:2019vub}. Today also NNLO
corrections to the decay to massive $b$-quarks are
known~\cite{Bernreuther:2018ynm,Behring:2019oci}. Ref.~\cite{Primo:2018zby}
also presents the exact computation of the top-Yukawa correction to
$\hbb{}$ decay, which appears first at NNLO.
As far as the partial width $\Gam{}$ is concerned, QCD corrections are
sizeable at NLO and more moderate at NNLO, i.e. the inclusive NLO/LO
K-factor in the massless $b$-quark limit is about $K^{m_b=0}_{\rm
  NLO/LO}\simeq 1.2$, whereas the NNLO/NLO inclusive K-factor is about
$K^{m_b=0}_{\rm NNLO/NLO}\lesssim
1.04$, cf.~\cite{DelDuca:2015zqa}. However, higher-order QCD corrections to
some kinematical distributions are very
large~\cite{Ferrera:2017zex,Caola:2017xuq,Gauld:2019yng}, making the
accurate description of this decay quite challenging.

The main Higgs production processes, gluon-fusion (ggf), Vector boson
Fusion (VBF) and associated HW/HZ production are known to NNLO
accuracy, or better, see e.g.\ Ref.~\cite{deFlorian:2016spz}.
Other Higgs production modes are also expected to be known to NNLO
accuracy in the upcoming years.
In view of this, combining the best accuracy in production and decay
seems natural. This was done in specific cases, e.g. for HW/HZ in
Refs.~\cite{Ferrera:2017zex,Caola:2017xuq,Primo:2018zby,Gauld:2019yng}.
It is well-known that corrections to production and decay can be
discussed separately to a very good approximation. In fact, since
the Higgs boson is a scalar, there are no spin-correlation
effects. Furthermore, the fact that it is a colour singlet implies
that non-factorisable corrections between production and decay vanish
exactly at NLO. At NNLO, the exchange of two gluons between production
and decay is possible, but these contributions are
${\cal O}(\as^2 \Gamma_H/M_H)$ for sufficiently inclusive observables,
i.e. for observables that are not sensitive to soft gluon
emissions~\cite{Fadin:1993kt,Fadin:1993dz}.

The description of the $\hbb{}$ decay at NNLO matched to a parton
shower is the main focus of this paper.
While NNLO corrections to the decay were already studied,
higher-order parton shower effects can still have large effects on
kinematic distributions of phenomenological relevance, in particular
when exclusive fiducial cuts are applied. Moreover, including NNLO
corrections into a fully exclusive event generator is particularly
important for analysis performed by the experimental collaborations.

In this work we use the \MINLO{}
method~\cite{Hamilton:2012np,Hamilton:2012rf} to build a generator
(henceforth named \hbbgMINLO{}) that produces fully exclusive
events for the $\hbb{}$ decay. Subsequently the accuracy of these
events is upgraded to NNLO by means of a reweighting procedure. Due to the
considerations made above, it is possible to interface such decay events
with any Higgs boson production process and further match them to a
parton shower. We explain how we implement this combination of
production and decay events, and what restrictions need to be
imposed on a parton shower evolution such that the fixed-order
accuracy is preserved.

The paper is organised as follows.
In Sec.~\ref{sec:NNLOPS} we focus on the decay of the Higgs to $b$
quarks and we first explain how to achieve predictions that are NLO
accurate both for $\hbb{}$ and $\hbb{g}$ observables using
\MINLO{}. Since the environment is particularly simple in this case,
we additionally present a numerical verification of the \MINLO{}
accuracy by performing a check in the $\as\to{0}$ limit. We then
detail the reweighting procedure to obtain NNLO accuracy for the
$\hbb{}$ decay width.
In Sec.~\ref{sec:interface} we describe the combination of
\hbbgMINLO{} generated decay events with a generic Higgs production
process and the subsequent combination with a parton shower.
In Sec.~\ref{sec:pheno} we show two explicit examples where we combine
NNLO $\hbb{}$ decay events with the HZ Higgs production process, as
well as with the VBF Higgs production process. We discuss the
advantages and limitations of the approach presented in this
article, and, for $\HZ$ production, we also compare our results with
those presented in Ref.~\cite{Astill:2018ivh}. We summarise our
findings in Sec.~\ref{sec:conclu}.
%

\section{$\hbb{}$ at NNLOPS accuracy}
\label{sec:NNLOPS}

In this section we describe a construction of an event generator that
describes the Higgs decay into $b$-quarks at NNLO QCD and allows for
inclusion of parton shower (PS) effects.
This is achieved by first merging the \hbb{} and \hbb{g} generators
using the \MINLO{} procedure and then performing a reweighting of the
\hbbgMINLO{} events to the exact NNLO \hbb{} partial width.

We focus here on the steps necessary to reach NNLO QCD
accuracy.
As a prerequisite we extended \POWHEG{} to incorporate a process
without initial-state radiation, where QCD partons are present only in
the final state.
We leave the discussion of the combination of NNLO-reweighted events
with a generic Higgs production mode, as well as interface with PS for
later sections.

\subsection{Merging $\hbb{}$ and $\hbb{g}$ generators using \MINLO{}}
\label{sec:merge}

We start by considering the decay of the Higgs boson into a pair of
massless $b$-quarks accompanied by an additional gluon,
\begin{align}
  H(p_1)
  \longrightarrow
  b(q_1) + \bar{b}(q_2) + g(q_3).
\end{align}
Our first goal is to construct an event generator that treats the
\hbb{g} processes at NLO accuracy and at the same time allows for
integrating out the extra gluon, yielding an NLO accurate description
of the \hbb{} decay channel.
We follow the usual \MINLO{}
procedure~\cite{Hamilton:2012np,Hamilton:2012rf} that enables us to
simulate the \hbb{} and \hbb{g} processes with a single event
generator without introducing an additional merging scale.
The method uses an NLO fixed-order calculation of the \hbb{g} process
together with information encoded in the Sudakov form factor of a
resummed prediction for some infrared safe observable that vanishes
for Born level $\hbb{}$ events.
In this case, we use a resummed prediction for the three-jet
resolution parameter $\yres$.
The $\yres$ observable corresponds to a value of a $\ycut$ parameter
within a clustering algorithm that separates between two- and
three-jet configurations or, equivalently, a maximal value of $\ycut$
for which two jets are obtained.
For reasons that will be explained in this subsection, we use the
Cambridge algorithm~\cite{Dokshitzer:1997in,Bentvelsen:1998ug} as a
clustering algorithm for the \hbb{(g)} merging procedure.
Its definition relies on two variables
\begin{align}
  v_{ij} &= 2(1-\cos{\theta_{ij}}), &
  y_{ij} &= v_{ij}\cdot \frac{\textrm{min}\{E_i,E_j\}^2}{\sdec}, 
\end{align}
where $\sdec$ is the squared centre-of-mass energy, i.e. the mass of the
on-shell Higgs boson, $E_i$, $E_j$ and $\theta_{ij}$ denote energies
and the angle between particles $i$ and $j$ in the Higgs boson
rest-frame.
During the clustering sequence, pairs of particles $(i,j)$ are sorted
according to the ordering variable $v_{ij}$, and the pair with a
minimal value of $v_{ij}$ is selected.
The procedure checks whether a test variable $y_{ij}$ is smaller than
a parameter $\ycut$, and if this is the case the particles $i$ and $j$
are recombined into a pseudo-particle and the algorithm repeats the
procedure with all remaining \mbox{(pseudo-)particles}.
On the other hand, if $y_{ij} > \ycut$, the softer of the two
pseudo-particles is considered a jet and removed from the list of
pseudo-particles.
The clustering procedure stops with all remaining
\mbox{(pseudo-)particles} declared as jets.

The resolution parameter $\yres$ classifies the hardness of events, i.e.
$\hbb{}$ events with additional hard, large angle gluons yield a large
value of $\yres$, close to the kinematical upper bound of 1/3, while
events with soft or collinear gluons have $\yres \ll 1$, which leads
to large logarithms $L\equiv -\ln \yres$ in the differential cross
section. Events with \hbb{} Born kinematics have $\yres=0$.

In order to merge the $\hbb{}$ and $\hbb{g}$ calculations with the
\MINLO{} method, we need a resummed calculation for $\yres$ that
includes all terms of order $\as^2 L$~\cite{Hamilton:2012rf}.
A next-to-next-to-leading-logarithmic (NNLL) accurate resummation of
the $\yres$ observable has been performed in Ref.~\cite{Banfi:2016zlc}
and we use this result to extract the required Sudakov form factor
$\Delta(\yres)$.
We decided to use the Cambridge algorithm as it is the one for which
the NNLL resummation for $\yres$ has the simplest structure.
To reconstruct the Sudakov form factor, we use the formulae given in
App.~B of Ref.~\cite{Banfi:2014sua}, setting $a=2$, $b_{\ell}=0$ for
the pure radiator function, and use the required NNLL multiple
emission corrections outlined in Eq.~(4) of Ref.~\cite{Banfi:2016zlc}.
Further details about the implementation of the NNLL ingredients
within the \MINLO{} formalism are given in App.~\ref{app:resformulae}.

Given the Sudakov form factor, the usual \MINLO{} formula for the
\POWHEG{} $\bar B$ function~\cite{Frixione:2007vw} has the form
\begin{align}
  \bar{B}(\Phibbg) =&
  \as(q_t^2) \Delta^2(y_3)
  \lb B_{\hbb{g}} \cdot \lp 1 - 2 \Delta^{(1)}(y_3)\rp   + V_{\hbb{g}} \rb\nonumber \\
  +&  \int d\Phi_r\ \as(q_t^2) \Delta^2(y_3) R_{\hbb{g}} 
  \,,
  \label{eq:bbar}
\end{align}
where $\Phibbg$ is the phase space of the three-body decay,
$B_{\hbb{g}}, V_{\hbb{g}}$ and $R_{\hbb{g}}$ denote the Born, virtual
and real matrix elements respectively, $\Delta^{(1)}(y_3)$ is the
${\cal O}(\as)$ expansion of the Sudakov form factor $\Delta(y_3)$,
and $q_t^2=\yres \sdec$. We notice that, in the first line of
Eq.~\eqref{eq:bbar}, $\yres$ is computed on the $\Phibbg$ kinematics,
whereas in the second line of Eq.~\eqref{eq:bbar}, $y_3$ is computed
on the full $H\to 4$ partons phase space $(\Phibbg,\Phi_r)$, as
required in \MINLO{}. All the
powers of $\as$, including the additional one contained in $R$, $V$
and $\Delta^{(1)}$, are evaluated at $q_t^2=\yres \sdec$, where
$\yres$ is defined on the appropriate phase space, i.e. $\Phibbg$ for
$V_{\hbb{g}}$ and $\Delta^{(1)}$, and $(\Phibbg,\Phi_r)$ for $R_{\hbb{g}}$.

As shown in Ref.~\cite{Hamilton:2012rf}, the integration over the
radiation phase space leads to an NLO accurate description of
$\hbb{}$ observables.
In particular, we obtain an NLO accurate result for the $\hbb{}$
partial width
\begin{equation}
  \GamMINLO{}
  \equiv \int d\GamMINLO{}= 
  \frac{1}{2 M_H} \int d \Phibbg \bar B(\Phibbg)\,. 
\end{equation}
A proof for the specific case at hand is
given in App.~\ref{app:analyticproof}.

We stress that this particular process offers a unique opportunity to
verify the absence of spurious ${\cal O}(\as^{3/2})$ terms in a simple
way.
Since no initial state partons are involved and we do not need to rely
on external input such as parton distribution functions, we can
easily consider the limit where $\as \to 0$ and check that the result
has the correct scaling, i.e.\ that the difference between $\GamMINLO$
and the nominal NLO result $\GamNLO$ is of ${\cal O}(\as^2)$.
We performed such a numerical check by manually changing the reference value of
the strong coupling, $\as(M_Z)$, at the level of \POWHEG{} input
card.\footnote{In order to reach the very small values of the strong
  coupling, about $\as(M_Z) \sim 0.02$, we needed to prepare and
  compile parts of the \POWHEG{} code in {\tt{quadruple}}
  precision. This is because the smaller $\as(M_Z)$ is, the more the
  Sudakov peak occurs at lower $\yres$ values and kinematical
  configurations with extremely small $\yres$ values are probed more
  frequently, spoiling the numerical stability.}
For the evaluation of the strong coupling at other scales, such as the
renormalisation scale $\mur$ or the resolution scale $q_t$, we use an
internal \POWHEG{} routine for the running of the strong coupling
implemented with the standard $\beta$ function.

\begin{figure}
  \centering 
  \includegraphics[width=0.49\linewidth]{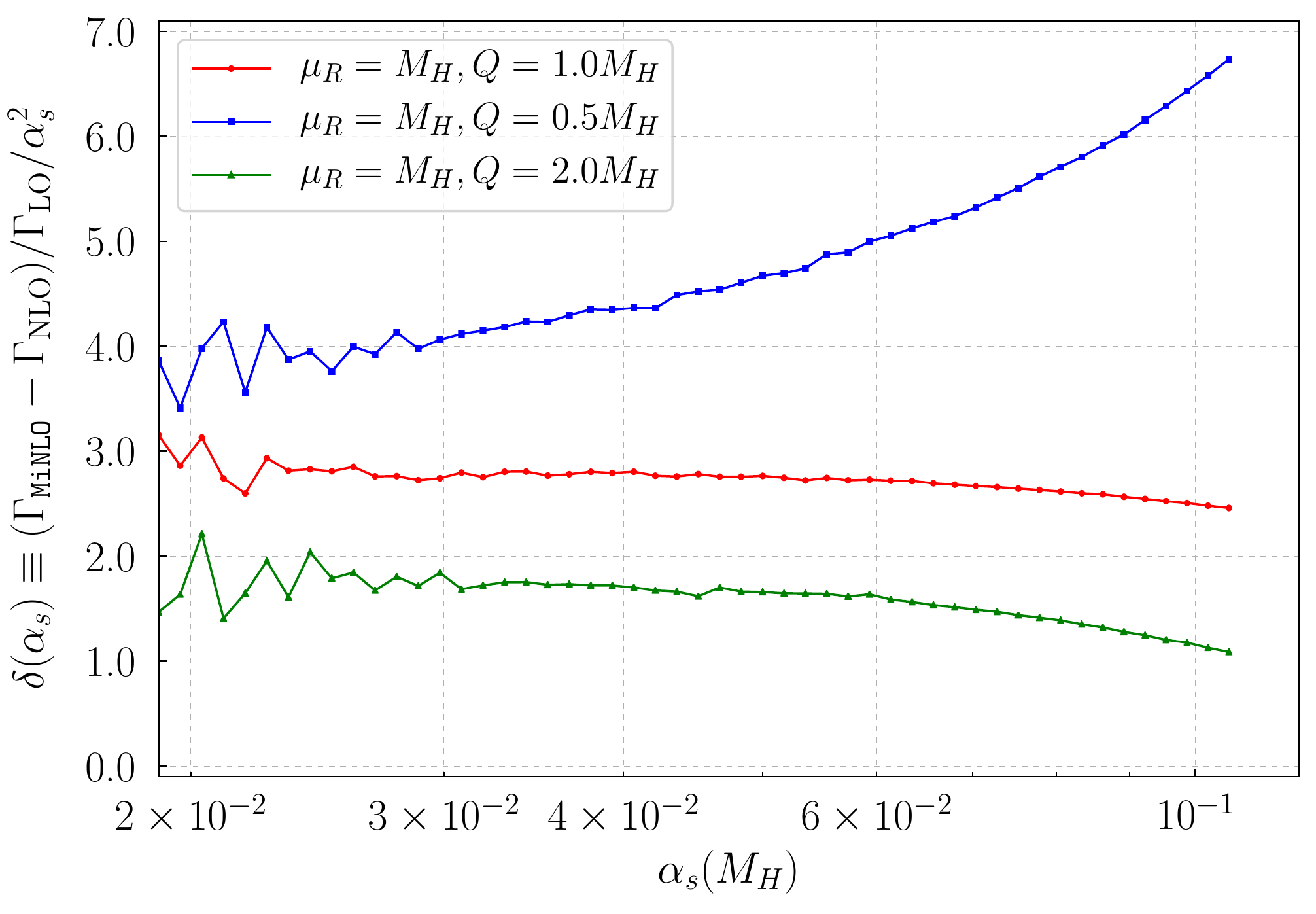}
  \includegraphics[width=0.49\linewidth]{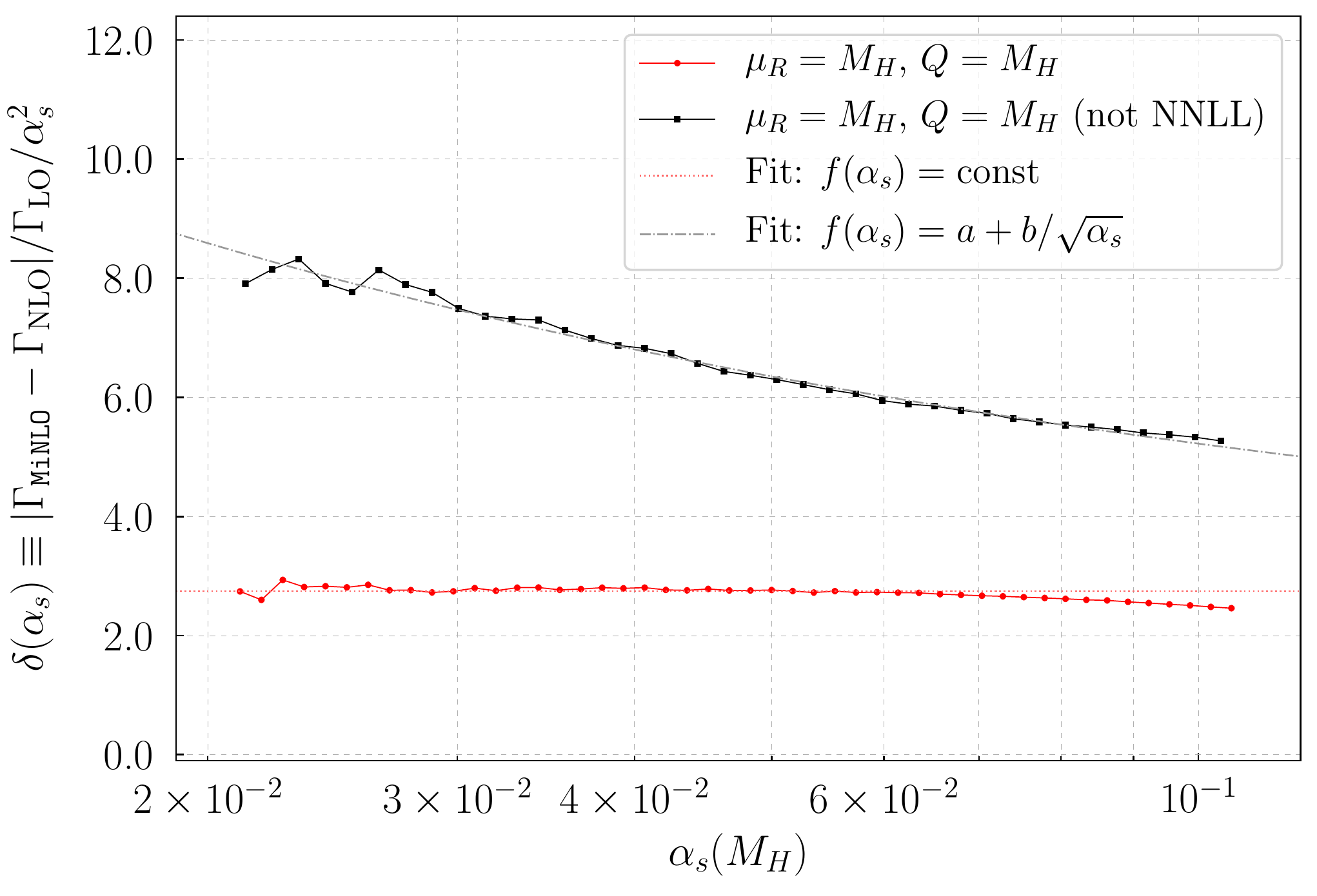}
  \caption{Numerical check of the NLO accuracy of the \hbbgMINLO{}
    result. The fact that $\delta(\as)$ approaches a constant at small
    $\as$, rather than increasing as $1/\sqrt{\as}$, shows that ${\cal O}({\as})$ terms in $\Gamma_{\MINLO}$ and $\Gamma_{\rm
      NLO}$ agree and that no spurious $\as^{3/2}$ terms are present in
    $\Gamma_{\MINLO}$.}
  \label{fig:numcheck}
\end{figure}
We present our results of the numerical check in
Fig.~\ref{fig:numcheck}.
We show the difference between the \hbbgMINLO{} result and the pure
NLO Higgs partial width to $b$-quarks.
We also normalise the difference to the LO width and further divide it
by a factor of $\as^2$. We write
\begin{equation}
  \label{eq:diff-minlo-nlo}
  \delta(\as) \equiv
  \frac{1}{\as^2} \cdot \frac{ \GamMINLO{} - \GamNLO{} }{ \GamLO{}}\,,
\end{equation}
where $\as$ is meant to be $\as(\mh)$.
We study the behaviour of $\delta(\as)$ as a function of the strong
coupling.

If spurious $\mathcal{O}(\as^{3/2})$ terms were present in the
\MINLO{} result, the difference defined in
Eq.~\eqref{eq:diff-minlo-nlo} would feature an increasing behaviour
when approaching the $\as \to 0$ limit.
On the contrary, we see that all curves in Fig.~\ref{fig:numcheck}
(left) approach a fixed value in the limit of small $\as$.
The three curves correspond to different values of the resummation
scale $Q$ introduced in App.~\ref{app:resformulae}
In Fig.~\ref{fig:numcheck} (right), we also show the difference
$\delta(\alpha_s)$ in the case when the NNLL resummation is not
included in the Sudakov form factor. The red curve is the one obtained
with the \hbbgMINLO{} generator, while the black curve is obtained
with the \hbbgMINLO{} generator with NNLL corrections related to
two-emission kernels removed, i.e. corrections included in
Eq.~\eqref{eq:fshift} were omitted. We see that this introduces the
spurious $\mathcal{O}(\as^{3/2})$ terms which are manifested as a
$1/\sqrt{\alpha_s}$ term in the difference $\delta(\alpha_s)$.

Before moving on to describe how we upgrade our results to reach NNLO
accuracy, we remind the reader that the function $\bar{B}(\Phibbg)$,
defined in Eq.~\eqref{eq:bbar}, is the weight used to generate points
in the $\Phibbg$ phase space according to the \MINLO{} method. The $\Phibbg$
phase space is the so-called \POWHEG{} ``underlying-Born'' phase space for the
case at hand. Therefore, once $\bar{B}(\Phibbg)$ is computed, the
fully differential partonic events are generated according the usual
\POWHEG{} procedure, i.e. the differential cross section reads:
\newcommand{\ptrad}{p_{t{\rm ,rad}}}
\newcommand{\LambdaPWG}{\Lambda_{\rm pwg}}
\begin{equation}
  d\sigma = d\Phibbg 
  \bar{B}(\Phibbg) \left[\Delta_{\rm pwg} (\LambdaPWG) +  \Delta_{\rm pwg} (\ptrad)  \frac{R_{\hbb{g}}(\Phibbg,\Phi_r)}{B_{\hbb{g}}(\Phibbg)} d\Phi_r
    \theta(\ptrad-\LambdaPWG)\right],
  \label{eq:fullpowheg}
\end{equation}
where
\begin{equation}
  \Delta_{\rm pwg} (p_t) = \exp{\left\{-\int \frac{R_{\hbb{g}}(\Phibbg,\Phi'_r)}{B_{\hbb{g}}(\Phibbg)} d\Phi'_r
    \theta(\ptrad' - p_t)\right\}}
\end{equation}
is the \POWHEG{} Sudakov form factor, $\LambdaPWG$ is an infrared
cutoff of the order of a typical hadronic scale, and $\ptrad$
corresponds to the transverse momentum of the secondary emission
associated with the radiation variables $\Phi_r$.  From
Eq.~\eqref{eq:fullpowheg} it is clear that the partonic events that
will be passed to the parton shower contain one or two extra
emissions. The NLO accuracy of $\hbb{g}$ observables is guaranteed by
the \POWHEG{} matching procedure, while the NLO accuracy for $\hbb{}$
observables is a consequence of the \MINLO{} method.

\subsection{Reweighting}
\label{sec:reweighting}

In order to achieve NNLO accuracy for $\hbb{}$-observables, we perform
a reweighting of \hbbgMINLO{} events with a reweighting factor
${\cal W}(\Phibb)$, where $\Phibb$ is the phase space of the
underlying $\hbb{}$ process.
In this particular case, because of the scalar nature of the Higgs
boson, the decay is isotropic in the Higgs boson rest frame and
therefore ${\cal W}(\Phibb)$ is just a number given by the constant
ratio
\begin{equation}
{\cal W} = \frac{\GamNNLO{}}{\GamMINLO{}}\,. 
\end{equation}
In order for the NNLO reweighting not to affect hard events, which are
already described at NLO accuracy by \MINLO{}, we modify the
reweighting factor as follows, cf.~\cite{Hamilton:2012rf},
\begin{equation}
{\cal W}(y_3) =  h(\yres)\cdot \frac{\GamNNLO{} -\GamMINLOb{}}{\GamMINLOa{}} +  (1-h(\yres))\,,   
\label{eq:Wy3}
\end{equation}
where
\begin{align}
\GamMINLOa{} &= \int d\Phi_{bbg} \lp \frac{d\GamMINLO{}(\Phi_{bbg})}{d\Phi_{bbg}} \cdot h(\yres) \rp  \,, \qquad \nonumber \\
\GamMINLOb{} &= \int d\Phi_{bbg} \lp \frac{d\GamMINLO{}(\Phi_{bbg})}{d\Phi_{bbg}} \cdot (1-h(\yres))\rp \,,
\end{align}
and the function $h(\yres)$,  
\begin{align}
  h(\yres) = \frac{ 1 }{ 1 +
    \lp\tfrac{\yres}{y_{3,\rm{ref}}}\rp^{y_{3,\rm{pow}}} }\,,
\end{align}
is used to classify the hardness of an event, i.e.\ $h(\yres)$
approaches one when the $b \bar{b}$-pair is accompanied by soft or
collinear emissions, while it should satisfy $h(\yres) \ll1$ for hard
events.
The value of $y_{3,\rm{ref}}$ should be choosen in such a way that the
weights of hard events are not significantly modified, hence it should
be rather small, but it should be above the Sudakov peak, such that
the NNLO reweighting is distributed among a large fraction of all
events.
Note that the three-jet resolution parameter cannot exceed $\yres =
1/3$ in three-body decays. Furthermore, in case of the \hbb{} decay,
the Sudakov peak is located at $\yres \approx e^{-5.5}$.
Taking these constraints into consideration, we choose $y_{3,\rm{pow}}
= 2$ and $y_{3,\rm{ref}} = e^{-4}$: this value corresponds to an event
with relative transverse momentum of about $17~\gev$, which can be considered a relatively
hard kinematics in the context of the \hbb{g} decay.

\section{Production-decay interface and parton shower matching}
\label{sec:interface}

In this section we describe how to combine the NNLO-reweighted events
produced with the \hbbgMINLO{} code with a generic Higgs production
process.
We also discuss how to deal with small off-shell effects.
Furthermore, we briefly describe how such hybrid events can be
consistently interfaced with a parton shower keeping the fixed-order
accuracy of the event sample, both in production and decay. Detailed
instructions and practical aspects are given in the manual
accompanying the public code.

We stress that the reweighting to NNLO, both in production and/or
decay, is optional, and if these steps are skipped one still obtains
NLOPS accuracy in production and decay.
For instance, one might consider including NNLOPS in decay, which is
numerically feasible, while keeping only NLOPS accuracy in the
production, in the cases where the multi-dimensional reweighting to get NNLOPS accuracy is numerically
very intensive or even unavailable, e.g. for VBF or $ttH$ processes.

\subsection{Combination of a generic Higgs production events with the \hbb{} events}
\label{sec:prod-dec-interface}
The starting point is to assume that one has access to a set of Les
Houches events produced with \POWHEGBOXVTWO{} with an undecayed Higgs
boson and possibly other final state particles.
Furthermore, one also has available a set of events with a Higgs-boson
decay into $b$-quarks plus additional QCD radiation prepared using the
\hbbgMINLO{} code, as described in the previous section.
In the following we describe the sequence of steps needed to
obtain a combined event that involves the production and the decay of a
Higgs boson in a melded event.

\paragraph{Momenta:}
Let us consider a single event record \evHprod{} with a Higgs boson in
the final state and possibly other final-state particles.
We denote the Higgs boson momentum by $\ph$.
On the other hand we have an event record \evHdec{} that describes a
decay of an on-shell Higgs boson, in its rest frame, to $b$-quarks
accompanied by additional radiation.
From these two sets of momenta we generate the full event as follows:
\begin{itemize}
\item We pull out the momenta of the Higgs decay products $q_i$ from
  the \evHdec{} record. Note that momenta $q_i$ are specified in the
  Higgs boson rest-frame.
\item From the momentum of the Higgs boson in the production stage,
  $p_H$, we compute a factor $\lambda = \sqrt{p_H^2}/M_H$.
  For an on-shell Higgs boson $\lambda = 1$, but if the Higgs boson in
  the {\evHprod} was generated with its virtuality distributed
  according to a Breit-Wigner shape then $\lambda \neq 1$.
\item In case $\lambda \neq 1$, we reshuffle the $q_i$ momenta such
  that the energy-momentum conservation is restored, i.e.  $q'_i =
  \mathbf{S}(\lambda) q_i$, such that $\sum_i q'_i = \lambda
  \mh$.\footnote{In this case the reshuffling is particularly simple and
    it amounts to rescaling the spatial momenta $\vec{q}_i$ by a common
    factor such that the energy-conservation is fulfilled.}
\item We boost momenta $q'_i$ from the Higgs boson rest frame to the
  laboratory frame, $p_i = \mathbf{B}(p_H) q'_i$, so that
  $\sum_i p_i = p_H$.
\item The final event record {\evHfull} is obtained by copying momenta
  and colour connections of all particles from the {\evHprod},
  changing the status of the Higgs boson to be an intermediate
  resonance ({\tt{istup = 2}}), and appending the $p_i$ momenta of the
  decay products with their colour connections at the end of the list.
\end{itemize}
Note that such a treatment of small off-shell effect neglects terms
suppressed by $(\Gamma_H/M_H)$.

\paragraph{Weights:}
Each production and decay event comes with its own set of weights due
to scale variations, i.e.~$\{w_{{\tt{prod}},i}\}$ and
$\{w_{{\tt{dec}},j}\}$.
The weights of the final {\tt{event\_Hprod\_x\_dec}} are obtained by
multiplying $w_{{\tt{prod}},i} \times w_{{\tt{dec}},j}$.
The most general set of weights is obtained by keeping all possible
$(i,j)$ combinations.
Nevertheless, simpler options are possible to consider, for instance,
by correlating the production and decay scale variations, as done for
the results presented in this paper.

\paragraph{PS radiation bound:}
As usual, \POWHEGBOX{} events come with an upper bound for additional
radiation that is generated during parton shower evolution, the
{\tt{scalup}} variable.
In the final {\evHfull} record we keep the original {\tt{scalup}} from
the production stage, i.e. {\scprod}, while the veto scale for the
decay, {\scdec}, can be easily reconstructed using the decay
kinematics.
We leave the discussion of further details to the next subsection.

\subsection{Interface to parton shower}
\label{sec:ps-interface}
We now discuss how to interface the final {\evHfull} record with a
parton shower. First of all, we note that many parton shower
generators enable inclusion of matrix element corrections to the
\hbb{} decay.  Nevertheless, the decay events produced with the
\hbbgMINLO{} code are already equipped with information encoded in
matrix elements for the \hbb{} decay with up to two additional
partons. Therefore, we make sure that the automatic matrix element
corrections of the parton shower program are switched off.

Furthermore, a consistent matching of an (N)NLO-accurate event
generator with a parton shower requires a special treatment of the
parton shower evolution.
Within the \POWHEGBOX{} framework, the hardest emission has already
occurred at the level of event generation and each event ({\evHprod} as
well as {\evHdec}, in our case) is equipped with an upper bound for
the subsequent radiation generated by a parton shower. This quantity
is commonly known as the {\tt{scalup}} variable.

As explained in the previous subsection, the full event record,
{\evHfull}, contains only the {\scprod} value. Nonetheless the bound
{\scdec} related to the \hbb{} decay stage can be reconstructed from
the decay kinematics using the \POWHEG{} definition of hardness in
final-state radiation~\cite{Campbell:2014kua}.
In particular, we identify the hardest splitting that occurred in the
decay using the information contained in the colour connections. We
denote by $p_{\rm rad}$ the radiated parton and by $p_{\rm em}$ the
parton that has emitted it, and then calculate the hardness of the
splitting $t$ as
\begin{equation}
  t = 2 p_{\rm rad} \cdot p_{\rm em} \frac{E_{\rm rad}}{E_{\rm em}}\,,  
  \label{eq:scalupdec}
\end{equation}
where $E_{\rm rad}$ and $E_{\rm em}$ are the energies of the two
partons in the Higgs boson rest frame.
We finally set {\scdec} to be $t$, whereas if the decay event has no
additional radiation besides the ``Born'' gluon in $\Phi_{bbg}$, we
set {\scdec} to be an infrared cutoff of
${\cal O}(\Lambda_{\rm QCD})$.

Such a treatment of the full event {\evHfull} provides a hardest
emission for the production stage and one for the decay stage.
This setup corresponds to the {\tt{allrad}} option described in App.~A
of Ref.~\cite{Campbell:2014kua}, hence, we follow the procedure
introduced in Ref.~\cite{Campbell:2014kua} to shower {\evHfull} events
with the \PYTHIA{8} parton shower.
More specifically, in order to respect the radiation bound coming from
the production stage we simply pass the {\tt{scalup}} variable, equal
to {\scprod}, to the shower program, as is always done when
interfacing \POWHEG{} with \PYTHIA{}.
In order to constrain the radiation emitted off the decay products
through showering, we implement a vetoed shower, i.e. we let
\PYTHIA{8} generate emissions off the decay products in all the
accessible phase space and, after the shower is completed, we check
the hardness of the splittings that were generated.
First, we find the splittings that originated from all the Higgs decay
products generated by \POWHEG{}, then, for each of them, we calculate
the corresponding hardness {\tt{hardness\_dec}} using
Eq.~\eqref{eq:scalupdec}.
If for each of those splittings, the corresponding hardness is smaller
than the veto scale {\scdec}, we accept the shower history; otherwise
we reject it and we attempt to shower the event again, until the
condition ${\tt{hardness\_dec}} < {\scdec}$ is met.
A very similar result can be obtained by using built-in facilities of
\PYTHIA{8}, namely by using the {\tt{UserHooks}}
class~\cite{Pythia8UserHooks,Pythia8POWHEG}.\footnote{We also notice
  that interfaces to shower \POWHEG{} events with multiple veto scales
  with \PYTHIA{8} and \noun{Herwig7} have been developed in
  Refs.~\cite{Ravasio:2018lzi,FerrarioRavasio:2019vmq}.}

\section{Practical implementation and phenomenological studies}
\label{sec:pheno}

In this section we present two concrete applications of the method
described in this article, which have NNLOPS accuracy in the \hbb{} decay.
We first consider the associated Higgs production process $pp \to H
Z(\to \ell \ell)$ described here at NNLOPS accuracy both in production
and decay, and compare results obtained here with those of
Ref.~\cite{Astill:2018ivh}, which have the same NNLOPS accuracy for
production but only NLOPS accuracy for the decay.\footnote{We note
  that {\tt Herwig++} code also allows to include NLO corrections to the \hbb{} decay in HV production~\cite{Richardson:2012bn}.}
The two results also differ in the treatment of the interface to the parton
shower in the decay, as explained in detail below.
Next, we consider vector boson fusion Higgs production, described at
NLOPS accuracy for the production stage. We interface the production with
NNLOPS accurate \hbb{} decay and we compare these results with those obtained by letting \PYTHIA{8} handle the Higgs decay.

\subsection{Associated Higgs production}

\paragraph{Input parameters and fiducial cuts:}
\label{sec:input}

As far as the Higgs production process is concerned, we used a setup
identical to Ref.~\cite{Astill:2018ivh} which facilitates a direct
comparison to those results.
In particular, we consider 13 TeV LHC collisions and use
\texttt{PDF4LHC15\_nnlo\_mc} parton distribution
functions~\cite{Ball:2014uwa,Harland-Lang:2014zoa,Dulat:2015mca,Carrazza:2015hva}.
We set $\mh = 125.0$ GeV, $\Gamma_H = 4.14$ MeV, $\mz=91.1876$ GeV, $\Gamma_Z = 2.4952$ GeV, $\mw=80.398$ GeV and $\Gamma_W = 2.141$ GeV. Moreover we set  
$G_F = 1.166387 \cdot 10^{-5} \gev^{-2}$, 
$\sin^2\theta_{W} = 0.23102$, 
and $\alpha_{\scalebox{0.5}{\rm{EM}}}(\mz) = 1/128.39$. 
The \hbb{} branching ratio is set to
$\BrHbb{} = 0.5824$~\cite{deFlorian:2016spz}.
For the contributions where the Higgs boson is radiated from a
heavy-quark loop we use the pole mass of the heavy quark. We set the pole
mass of the bottom quark to $m_b = 4.92~\gev$ and the pole mass of the
top quark to $m_t = 173.2~\gev$.

For the decay, our implementation of \hbbgMINLO{} uses the matrix
elements from Ref.~\cite{DelDuca:2015zqa}.
The bottom Yukawa coupling in $\hbb{}$ decay is evaluated in the
$\MSbar{}$ scheme at the decay renormalisation scale $\murdec =
\mh$.
The running bottom Yukawa coupling is computed from the on-shell
Yukawa coupling using an ${\cal O}(\alpha_s^2)$ conversion that is
implemented along the lines of {\tt{RunDec}}
package~\cite{Chetyrkin:2000yt,Herren:2017osy}.
The numerical value of the bottom quark $\MSbar$ mass at the
renormalisation scale is $\mbMSbar(\mh) = 3.152~\gev$ and the
corresponding Yukawa coupling is $y_b(\mh) = 1.280 \cdot 10^{-2}$.
Note that at variance to the Ref.~\cite{Astill:2018ivh} the $b$-quarks
are treated as massless inside the \hbbgMINLO{} generator.

The production events were reweighted to the NNLO accuracy using NNLO
fixed-order predictions from \MCFM{}~\cite{Campbell:2016jau} with
central factorisation and renormalisation scales set to the sum of the
Higgs boson and the Z boson mass, $\mu = \mh+\mz$.
The NNLO prediction for the decay was obtained from an analytical
result of Ref.~\cite{Chetyrkin:1996sr} and uses the Higgs boson mass
as a central renormalisation scale.
When interfacing the fixed-order predictions with a parton shower we
use \PYTHIA{8}~\cite{Sjostrand:2014zea} using the Monash
tune~\cite{Skands:2014pea}.

The scale uncertainty is obtained by correlating the scale variation
factors $(K_r,K_f)$ of the \MINLO{} and NNLO predictions that enter
the reweighting procedure.
The theoretical uncertainty is estimated by performing the usual 7
point scale variation, i.e. it corresponds to an envelop of 7 scale
combinations, according to $1/2 \le K_r/K_f < 2$.

We consider the same fiducial cuts as in
Ref.~\cite{Astill:2018ivh}. In particular, we require two charged
leptons with $|y_{\ell}| < 2.5$ and $p_{t,\ell} > $ 7 GeV. The harder
lepton should satisfy $p_{t,\ell} > $ 27 GeV, the invariant mass of
the leptons-system should be in the range 81 GeV $ < \mll < $ 101 GeV.
Additionally we require at least two $b$-jets with $|\eta_{b}| < 2.5$
and $p_{t,b} > $ 20 GeV.  Jets are defined using the flavour-$k_t$
algorithm~\cite{Banfi:2006hf}, where we consider only
$b$-quarks to be flavoured, and all other light quarks to be
flavourless.

\begin{figure}
\centering 
\includegraphics[width=0.48\linewidth]{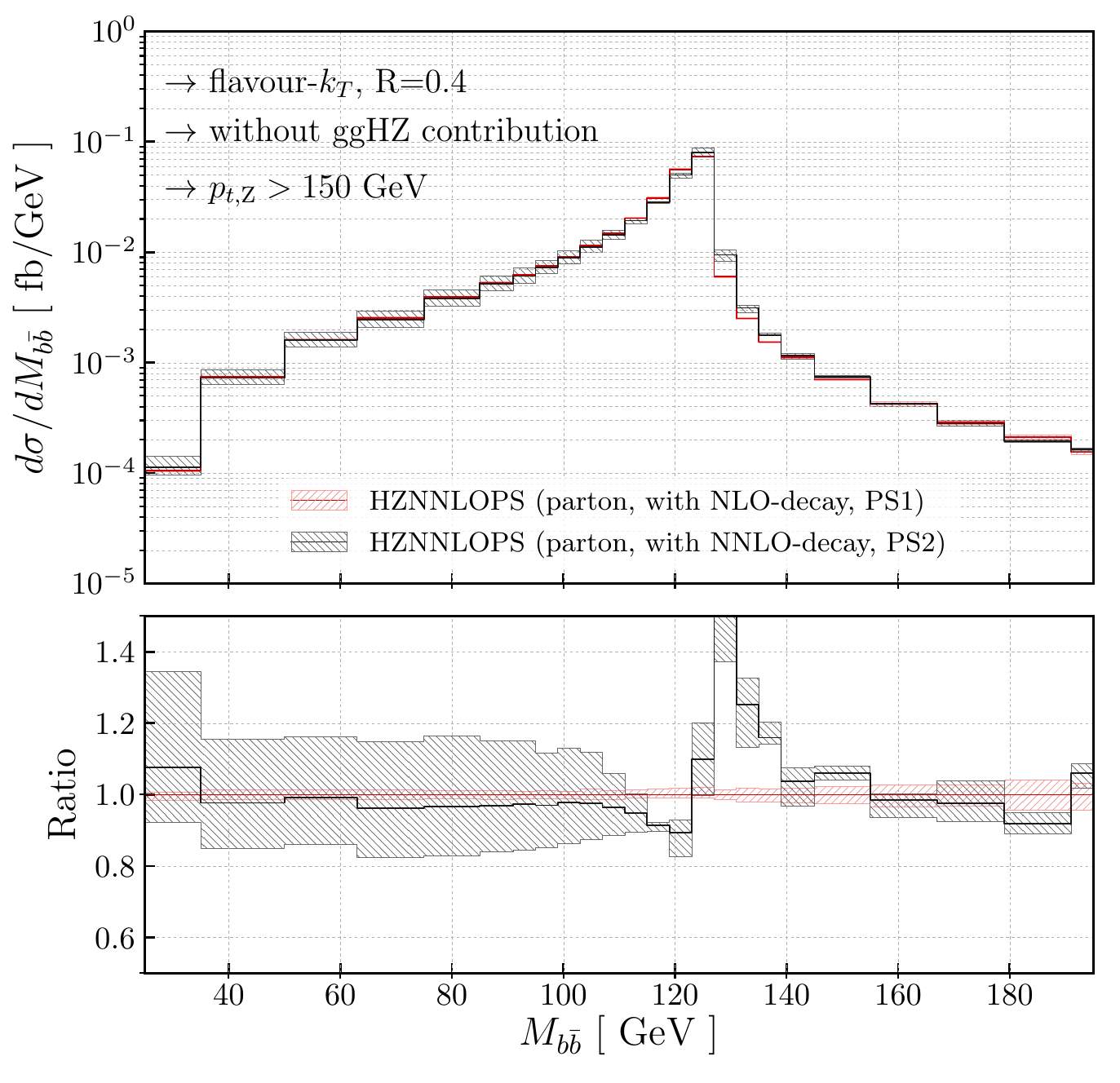}\hfill 
\includegraphics[width=0.48\linewidth]{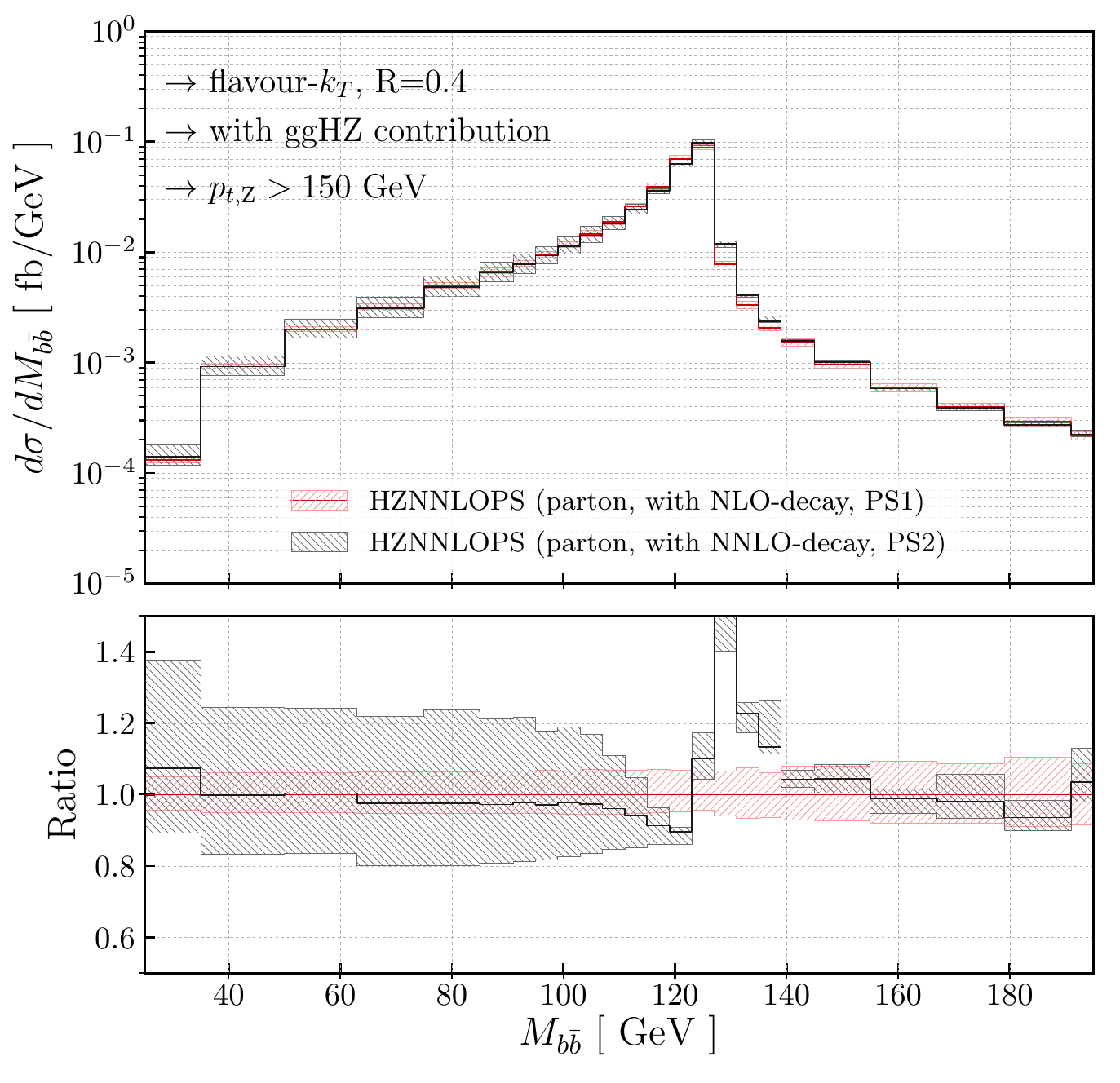}
\includegraphics[width=0.48\linewidth]{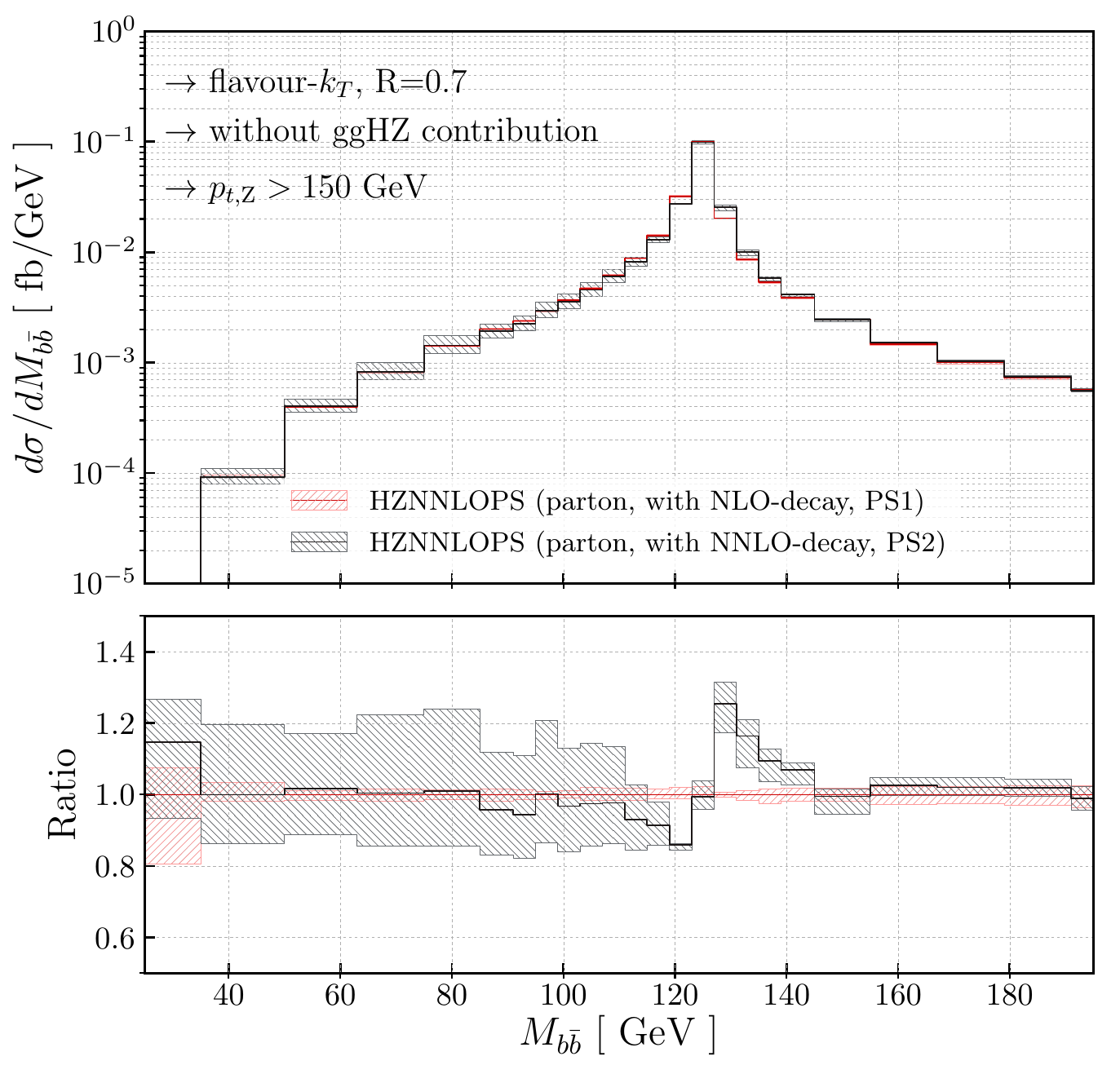}\hfill
\includegraphics[width=0.48\linewidth]{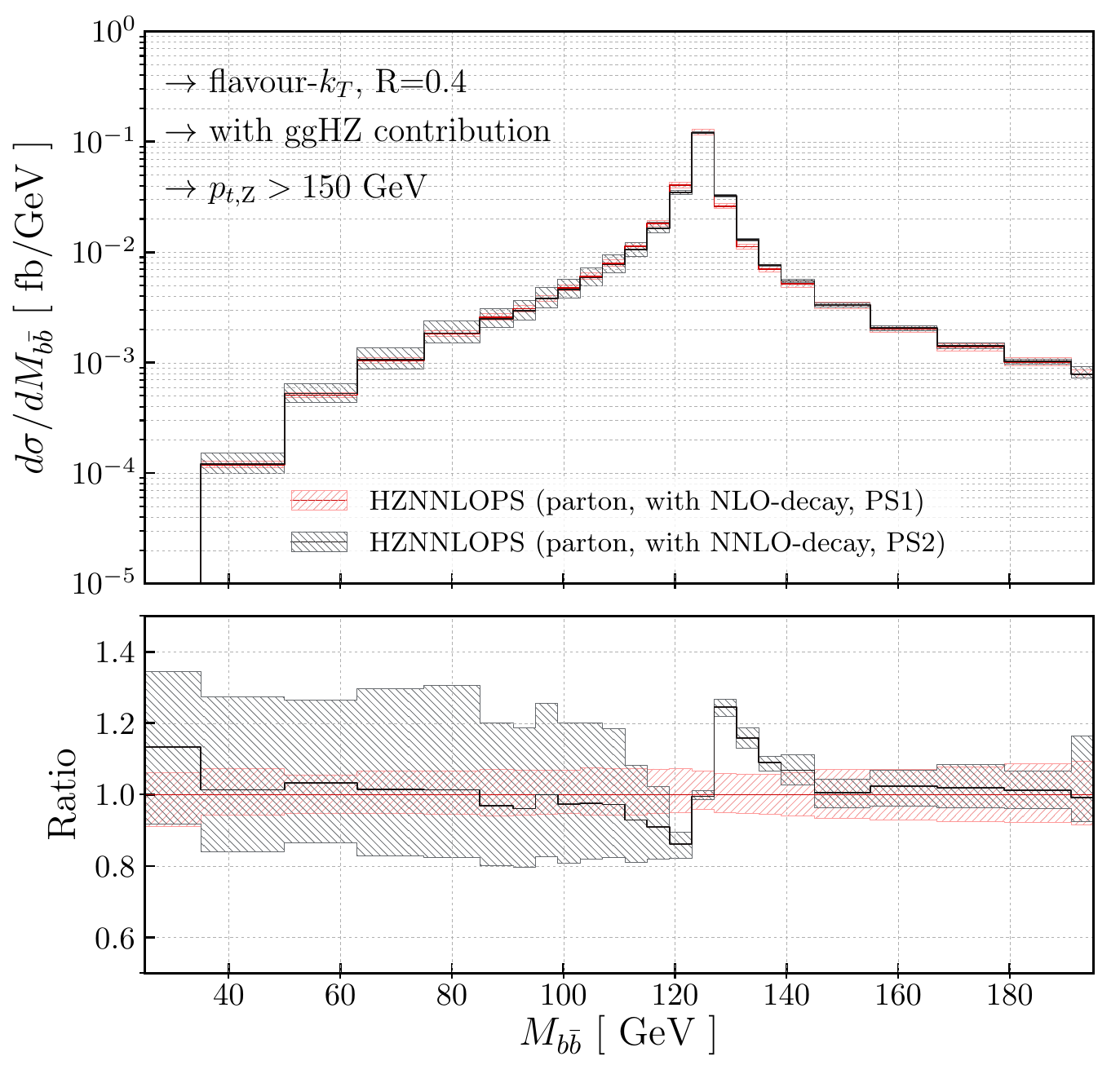}
\caption{Invariant mass of the two $b$-jets reconstructed using with
  the flavour-$k_t$ algorithm with $R=0.4$ (upper plots) and $R=0.7$
  (lower plots). Left (right) plots are without (with) gluon-induced
  terms.  We show the NNLOPS predictions of Ref.~\cite{Astill:2018ivh}, which
  have only NLO corrections to the decay (red) and the predictions of
  this work (black). Predictions of Ref.~\cite{Astill:2018ivh} and
  this work also differ in the handling of the matching to the parton
  shower, see text for more details.}
  \label{fig:mbb}
\end{figure}

\paragraph{Results:}
We start by showing the invariant mass of the two $b$-jets in
Fig.~\ref{fig:mbb}. The jets are reconstructed using the flavour-$k_t$
algorithm~\cite{Banfi:2006hf} with $R=0.4$ (upper plots) and $R=0.7$
(lower plots).\footnote{We use the flavour-$k_t$ algorithm to be
  aligned with the analysis presented in
  Ref.~\cite{Astill:2018ivh}. We also note that, since $b$-quarks are
  given a mass by \PYTHIA{8}, other algorithms can be used, however,
  from a theoretical point of view, the flavour-$k_t$ algorithm allows
  one to reduce the sensitivity of predictions on the value of the
  $b$-quark mass.}
If more $b$-jets are present in the final state, the pair with the
invariant mass closer to $\mh$ is selected.
Plots on the right-hand-side include loop-induced $gg \to HZ$
contribution.
The cuts applied here are those described earlier in this subsection
with an additional cut of $p_{t,{\rm Z}} = 150~\gev$ to select boosted
Higgs-decays.
The plots show the NNLOPS predictions
of Ref.~\cite{Astill:2018ivh}, which have NNLO accuracy in
production and only NLO corrections to the decay (red) and the
predictions of this work (black), which are NNLO accurate both in
production and decay.
The labels ``PS1'' and ``PS2'' make the reader aware of the fact that
differences between the two NNLOPS predictions are not only due to the
NLO vs NNLO treatment of the decay, but also to the different handling of
the matching to the parton shower, which for PS1 uses a single overall
{\tt{scalup}} bound for a whole event, as described in Sec.~3.2 of
Ref.~\cite{Astill:2018ivh}, while the PS2 method comes with a separate
{\tt{scalup}} bound for production and decay stages, as outlined in
Sec.~\ref{sec:ps-interface} of this work.
The results of Ref.~\cite{Astill:2018ivh} require only one
\texttt{scalup} variable since in that case the \POWHEG{} hardest
emission is generated at most from one singular region, associated
either with the production stage or with the radiation from the decay.
We remind the reader that for each event all singular regions are
probed and only the hardest one is kept in the record.\footnote{This
  setup corresponds to \texttt{allrad = 0} option of the App.~A of
  Ref.~\cite{Campbell:2014kua}.}

We first focus on the plots without gluon-induced contributions. The
most striking difference between the NLO-decay-PS1 (red) and
NNLO-decay-PS2 (black) predictions is in the size of the uncertainty
bands. As was already discussed in Ref.~\cite{Astill:2018ivh}, the
uncertainty of the NLO-decay-PS1 result is underestimated due to the
well-known feature of the \POWHEG{} simulation, i.e. the scale is
varied at the level of the $\bar B$ function, which is inclusive over
radiation, while the $\mbb$ spectrum is sensitive to secondary
radiation.
On the other hand the NNLO-decay-PS2 result has a more realistic
uncertainty estimate, as the uncertainty is driven by the
renormalisation scale variation in Eq.~\eqref{eq:bbar}.

Within their uncertainties, NLO-decay-PS1 and NNLO-decay-PS2
predictions agree very well in the whole $\mbb$ spectrum with the
exception of the region close to $\mbb=\mh$.
In particular, we notice that just above (below) the peak the
NLO-decay-PS1 prediction lies below (above) the NNLO-decay-PS2
prediction. These differences are not unexpected.
Notably, the hardest emission from the decay was treated differently
in the two results, i.e. in the NLO-decay-PS1 approach it was
generated using the \POWHEG{} Sudakov, while in the NNLO-decay-PS2
simulation it is controlled by the \MINLO{} Sudakov.
Furthermore, the composition of the Les Houches events in the two
cases is rather different. For instance, in the NLO-decay-PS1 approach
one can start with pure $\hbb{}$ decay events, not accompanied by
additional radiation, where subsequent emissions from the decay
generated completely by the parton shower, with a value of {\tt
  scalup} determined by the hardest \POWHEG{} emission in
production. In NNLO-decay-PS2 events instead one can have events with
two emissions from the decay, which can further be combined with a
production event involving up to two additional emissions.
While both methods are formally correct, numerical differences are not
surprising.
Altogether, the NNLO-decay-PS2 approach has the advantage that more
emissions are generated using exact matrix elements which are properly
matched to PS, whereas in the NLO-decay-PS1 approach part of this task
was left to the parton shower alone.

Similar effects are observed when the gluon-induced contribution is
included, as presented in the right-hand plots in Fig.~\ref{fig:mbb}.
In this case, an additional difference between the two approaches is due to
the fact that all the decays in the NNLO-decay-PS2 event-sample are
generated using the interface method described in
Sec.~\ref{sec:prod-dec-interface}, whereas in the case of
NLO-decay-PS1 the decay for these events, which are formally already
${\cal O}(\as^2)$, was always generated by \PYTHIA{8}.

\begin{figure}
\centering 
\includegraphics[width=0.48\linewidth]{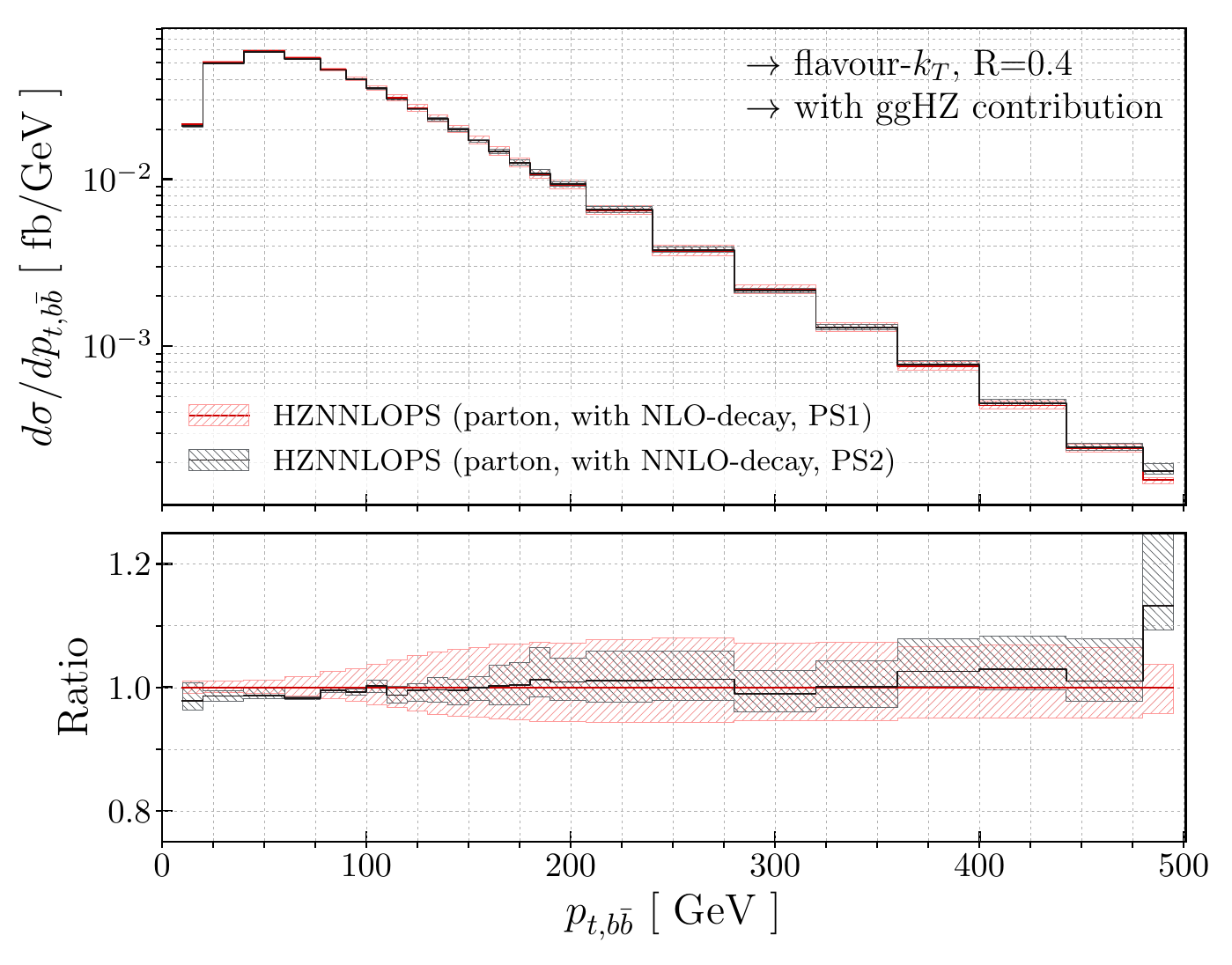}\hfill 
\includegraphics[width=0.48\linewidth]{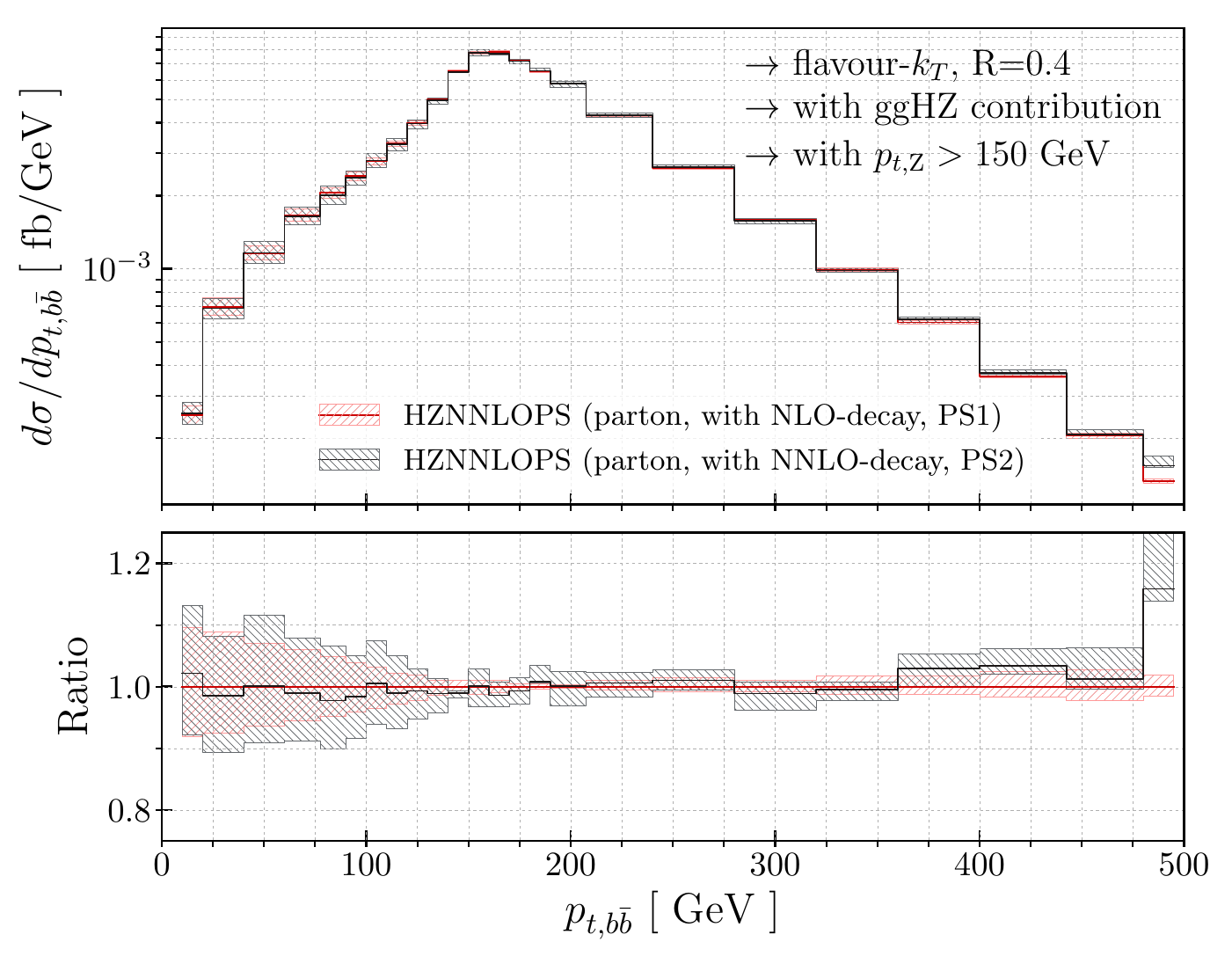}
\caption{As Fig.~\ref{fig:mbb} but for the transverse momentum of the
  two $b$-jets used for the Higgs reconstruction, $p_{t, b\bar b}$,
  without (left) and with (right) a $p_{t, Z}$ cut of 150 GeV.}
\label{fig:ptbb}
\end{figure}
Next, in Fig.~\ref{fig:ptbb} we show the transverse momentum of the
two $b$-jets used for the Higgs reconstruction, $p_{t, b\bar b}$,
without (left) and with (right) a $p_{t, Z}$ cut of 150 GeV.
Again the main features of these plots were already described in
Ref.~\cite{Astill:2018ivh}, see discussion of Fig.~9 therein.
Here, we focus on the comparison between the predictions of that paper
and this work. We note that, in this case, both the NLO-decay-PS1 and
the NNLO-decay-PS2 predictions have realistic scale-uncertainty
bands. The uncertainty band of the latter result is slightly
narrower. This is expected since, for the new result, NNLO corrections
in the decay are included.

\begin{figure}
\centering 
\includegraphics[width=0.48\linewidth]{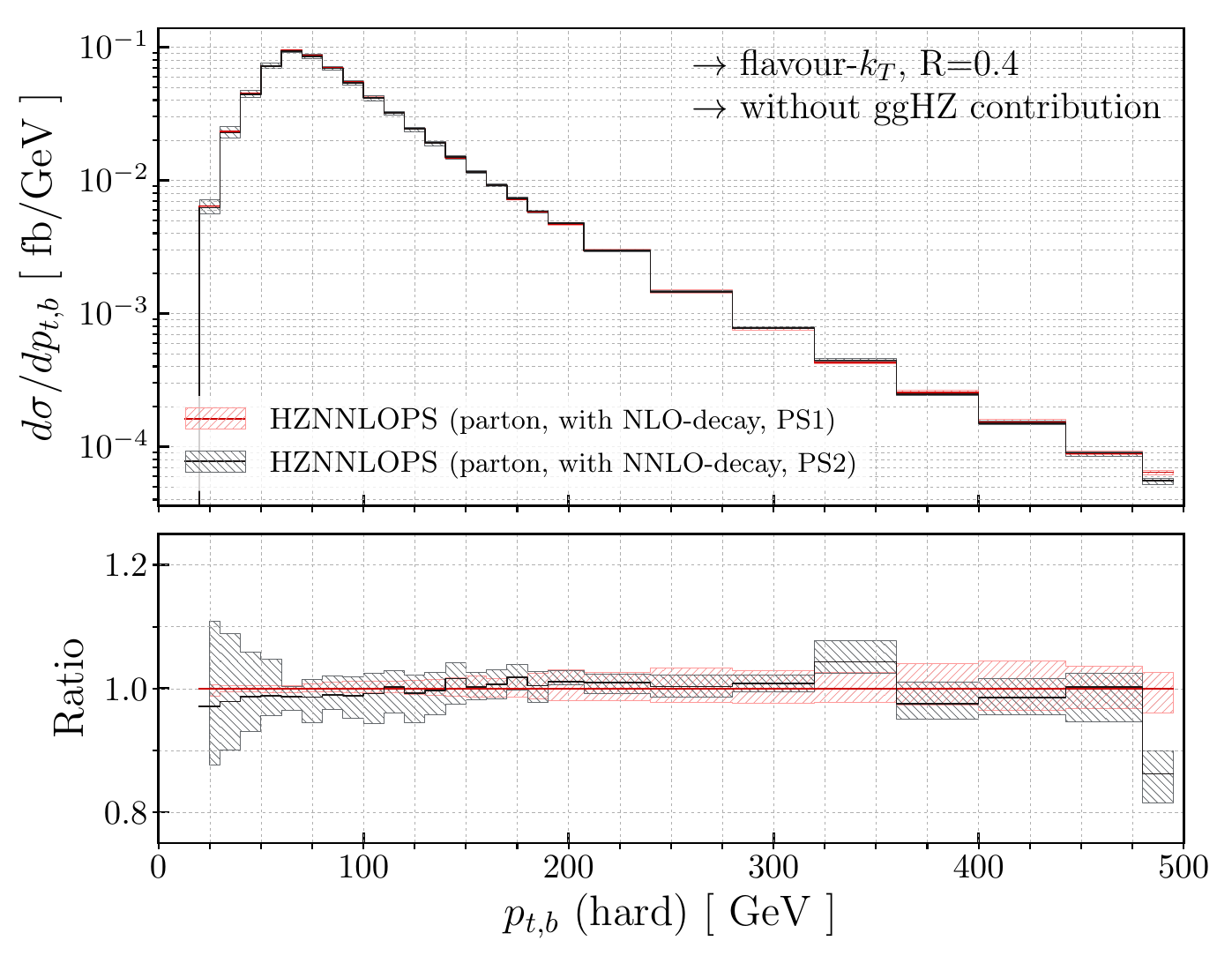}\hfill 
\includegraphics[width=0.48\linewidth]{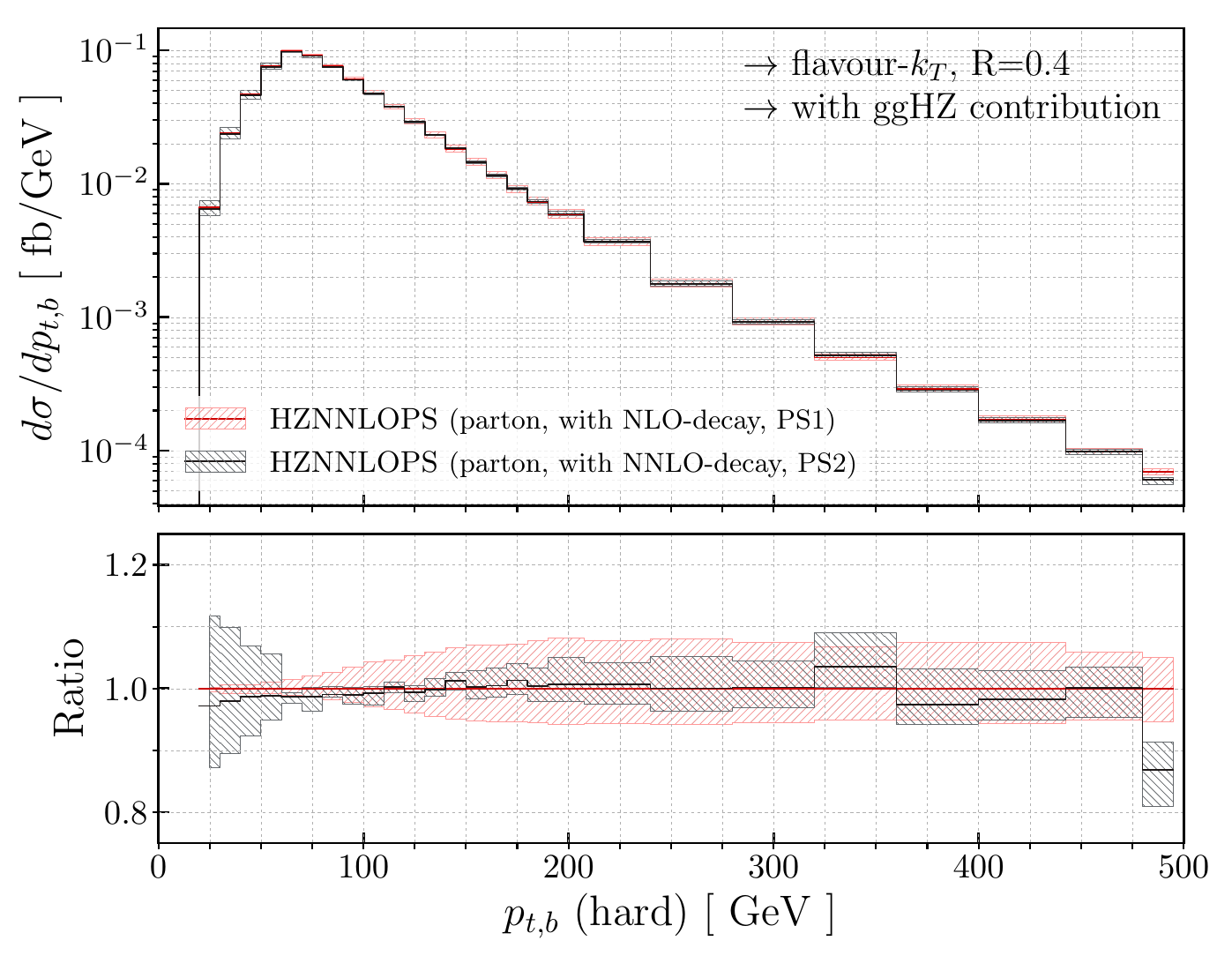}
\caption{As Fig.~\ref{fig:mbb} but for the hardest of the two $b$-jets
  used for the Higgs reconstruction, $p_{t, b}$(hard), without (left)
  and with (right) the gluon-induced terms, and without the cut on $p_{t,Z}$.}
\label{fig:ptb}
\end{figure}
In Fig.~\ref{fig:ptb} we show the transverse momentum of the hardest of the two $b$-jets,
$p_{t, b}$(hard).
Both plots are considered without the boosted Higgs-decay selection,
i.e. the $p_{t, Z} > 150~\gev$ cut is omitted.
The left (right) plot is without (with) gluon-induced terms.
In the right-hand side plot, we also notice that, when the
gluon-induced terms are included, the uncertainty bands of all
predictions gets larger, as can be seen clearly above $\sim 100$~GeV.
As was already discussed in
Ref.~\cite{Astill:2018ivh}, due to their kinematics, this is the
region where these events contribute, and the widening is due to the
fact that they have a LO-like uncertainty band.

Finally, in the region up to $\sim 80$~GeV, the NLO-decay-PS1
uncertainty band is considerably smaller than the one of the NNLO-decay-PS2
result. This region is populated mostly by events with at least one
relatively hard emission from the decay. When such an emission is
generated as the \POWHEG{} hardest emission (NLO-decay-PS1), the scale
uncertainty is underestimated (see discussion about
Fig.~\ref{fig:mbb}). The NNLO-decay-PS2 result provides a more
accurate uncertainty band.

\subsection{Vector boson fusion Higgs production}
\label{sec:vbf}

To demonstrate the flexibility of our program, we consider now the NLO
production of a Higgs boson via vector boson fusion (VBF) matched to
parton shower, as implemented in \POWHEG{}~\cite{Nason:2009ai}.
We consider the Higgs boson decaying to $b$-quarks, and study the
effect of including NNLOPS corrections to the decay.
We compared our predictions to a more standard treatment where
\PYTHIA{8} fully handles the Higgs decay and all subsequent radiation
off the decay products.

\paragraph{Input parameters and fiducial cuts:}
\label{sec:cutsvbf}
For VBF, we consider 13 TeV LHC collisions and use
\texttt{PDF4LHC15\_nlo\_mc} parton distribution functions. We set
$\mh = 125.0$ GeV, $\Gamma_H= 4.088$ MeV, $\mz=91.1876$ GeV,
$\Gamma_Z = 2.4952$ GeV, $\mw=80.398$ GeV and $\Gamma_W = 2.141$
GeV. Moreover, we use $\sin^2\theta_{W} = 0.2310$ and
$\alpha_{\scalebox{0.5}{\rm{EM}}} = 1/128.93$. The \hbb{} branching
ratio is fixed to
$\BrHbb{} = 0.5824$~\cite{deFlorian:2016spz}.

The scale variation procedure we use for VBF is similar to the one
used for HZ events, except that here we further restrict to a 3-points
scale variation by correlating renormalisation and factorisation scale
variation, i.e. $K_r=K_f$. We also fully correlate the renormalisation
scale variation for the production and the decay.

In order to show results in a fiducial phase space which is similar to the one
probed in a real analysis, we apply cuts that enhance the vector boson
fusion signal with respect to its typical backgrounds. In particular
we require two tagging jets with $p_{\rm t,j} > 30$\,~GeV, $|y_{\rm
  j}| < 5$, $y_{j1}\cdot y_{j2}<0$, $M_{\rm jj} > 600$\,~GeV and $|y_{j1}-y_{j2}|> 4.2$.
Furthermore, we require the presence of two $b$-jets with
$p_{t,b} > 20$\,~GeV and $|y_{b}| < 2.5$.
As for the HZ results shown above, jets are defined using the
flavour-$k_t$ algorithm, where we consider only $b$-quarks to be
flavoured, and all other light quarks to be flavourless.

\begin{figure}
\centering 
\includegraphics[width=0.45\linewidth]{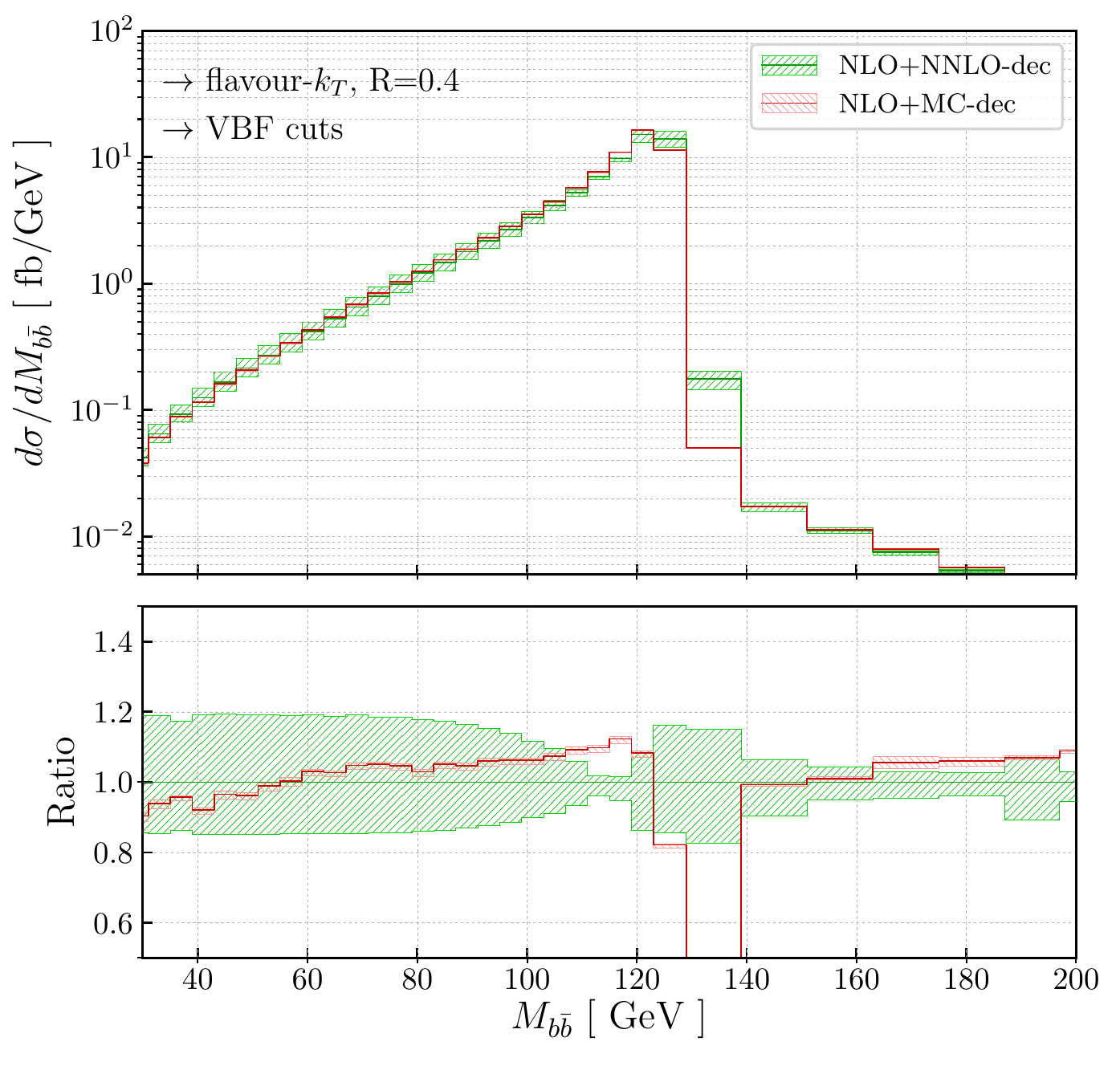}\hfill
\includegraphics[width=0.45\linewidth]{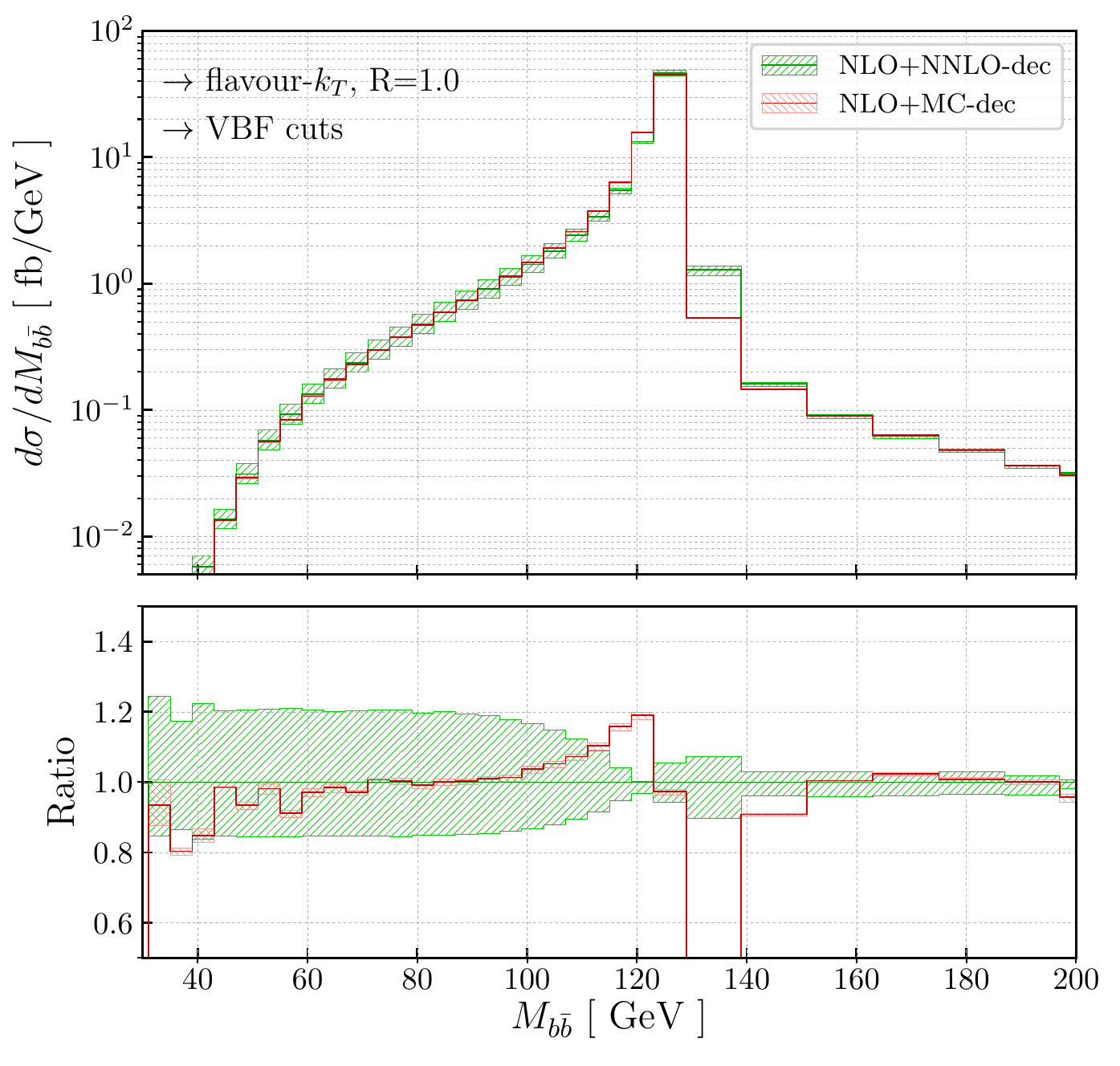}
\caption{Invariant mass of the two $b$-jets in VBF Higgs events with
  Higgs decaying to $b$-quarks. The VBF cuts described in the text are applied. The jets are reconstructed with the
  flavour algorithm of Ref.~\cite{Banfi:2006hf} for $R=0.4$ (left) and $R=1.0$
  (right).}
\label{fig:vbf_mbb}
\end{figure}

\paragraph{Results:}
As for the results shown in the previous section, the more interesting
observables to show the effect of the NNLOPS treatment of the \hbb{}
decay are those obtained from the $b$-jet kinematics, and we focus on
the invariant mass of the two $b$-jets ($\mbb$), shown in
Fig.~\ref{fig:vbf_mbb}, and the transverse momentum of the hardest of
the two $b$-jets ($p_{t, b}$(hard)) in Fig.~\ref{fig:vbf_ptb}.

Jets are reconstructed using the flavour-$k_t$
algorithm~\cite{Banfi:2006hf}, and we show results for $R=0.4$ (left
plots) and $R=1$ (right plots).
As done for HZ, if more than two $b$-jets are present in the final
state, the pair with the invariant mass closes to $\mh$ is selected.

For VBF, we compare our new result (NLOPS for VBF production, NNLOPS
for \hbb{}) against the result where the \hbb{} decay is handled by
\PYTHIA{8}. These results are labeled in the plots as
``NLO+NNLO-dec'' (green) and ``NLO+MC-dec'' (red), respectively.

The NNLOPS treatment of the decay has its most important impact on the
invariant mass of the two $b$-jets that originate from the Higgs
decay, shown in Fig.~\ref{fig:vbf_mbb}.
First we focus on the shapes of the central predictions.  First of
all, one can notice that, above the peak, the cross section drops very
quickly, much more than in the HZ case, see Fig.~\ref{fig:mbb}. This
is due to the nature of VBF production, in particular when tight VBF
cuts are applied. In fact, for $\mbb$, the region above the peak is
dominated by radiation from production being clustered together with
the $b$-jets, i.e. a configuration which is clearly suppressed by the
VBF cuts, where the light jets are required to be far from the central
rapidity region, where the Higgs boson is typically found.

Far from the peak, the two results are in rather good agreement, and
we observe a pattern not too different from what was observed in
Ref.~\cite{Astill:2018ivh}, see Fig.~10 therein. We attribute the
shape difference observed just below the peak ($100~\gev \lesssim \mbb
< \mh$) to differences between the \PYTHIA{8} and the \MINLO{}
Sudakov.
The distribution in the left plot of the Fig.~\ref{fig:vbf_mbb} peaks
at a slightly smaller value of $\mbb$ than it does in the right plot.
This behaviour is attributed to the jet radius used in the clustering
sequence, i.e. for a bigger jet radius ($R=1.0$, right plot) more
radiation is clustered within $b$-jets moving the peak to higher
$\mbb$ values than for smaller jet radius ($R=0.4$, left plot).
The differences close to the peak, driven by soft emissions off the
$b$-quarks, can be associated with the different treatment of their
masses in the two results, i.e. the ``NLO-NNLO-dec'' features massless
$b$-quarks in the \hbbgMINLO{} part of the simulation while the ``NLO-MC-dec'' one
treats the Higgs decay to massive $b$-quarks.
Note that the difference seems to be particularly large in the first
bin to the right of the peak. This is not surprising since the
absolute values of the cross-section in the neighbouring bins differ
by almost two orders of magnitude and even a small migration of events
from the peak induces large effects in the next bin. We study these
mass effects more thoroughly in App.~\ref{app:mb0}.

The size of the uncertainty band in Fig.~\ref{fig:vbf_mbb} also
requires some explanation. It is about 15-20\% in the case of the
``NLO+NNLO-dec'' prediction below the Higgs mass, and about 5\%
above. This has to be contrasted with an uncertainty band which is at
most a few percent throughout the whole mass spectrum when the decay
is handled by \PYTHIA{8}.
The scale uncertainty band for the ``NLO+NNLO-dec'' result is as
expected: below the peak the result features a scale-variation
uncertainty in line with the fact that this is the kinematic region
dominated by radiation off the $b$-quarks, and thereby the size of the
uncertainty is of NLO type. This is the same pattern as observed in
the HZ case.  In the ``NLO+MC-dec'' situation, the smallness of the
band is due to a well-known \POWHEG{} feature, namely the fact that
the only source of scale dependence comes from the \POWHEG{} $\bar{B}$
function, which is an inclusive quantity with respect to the radiation
kinematics. For VBF, the NLO scale dependence is very small, and since
we are not performing any scale variation within \PYTHIA{8}, the
uncertainty band for $\mbb$ is very small.

\begin{figure}
\centering 
\includegraphics[width=0.45\linewidth]{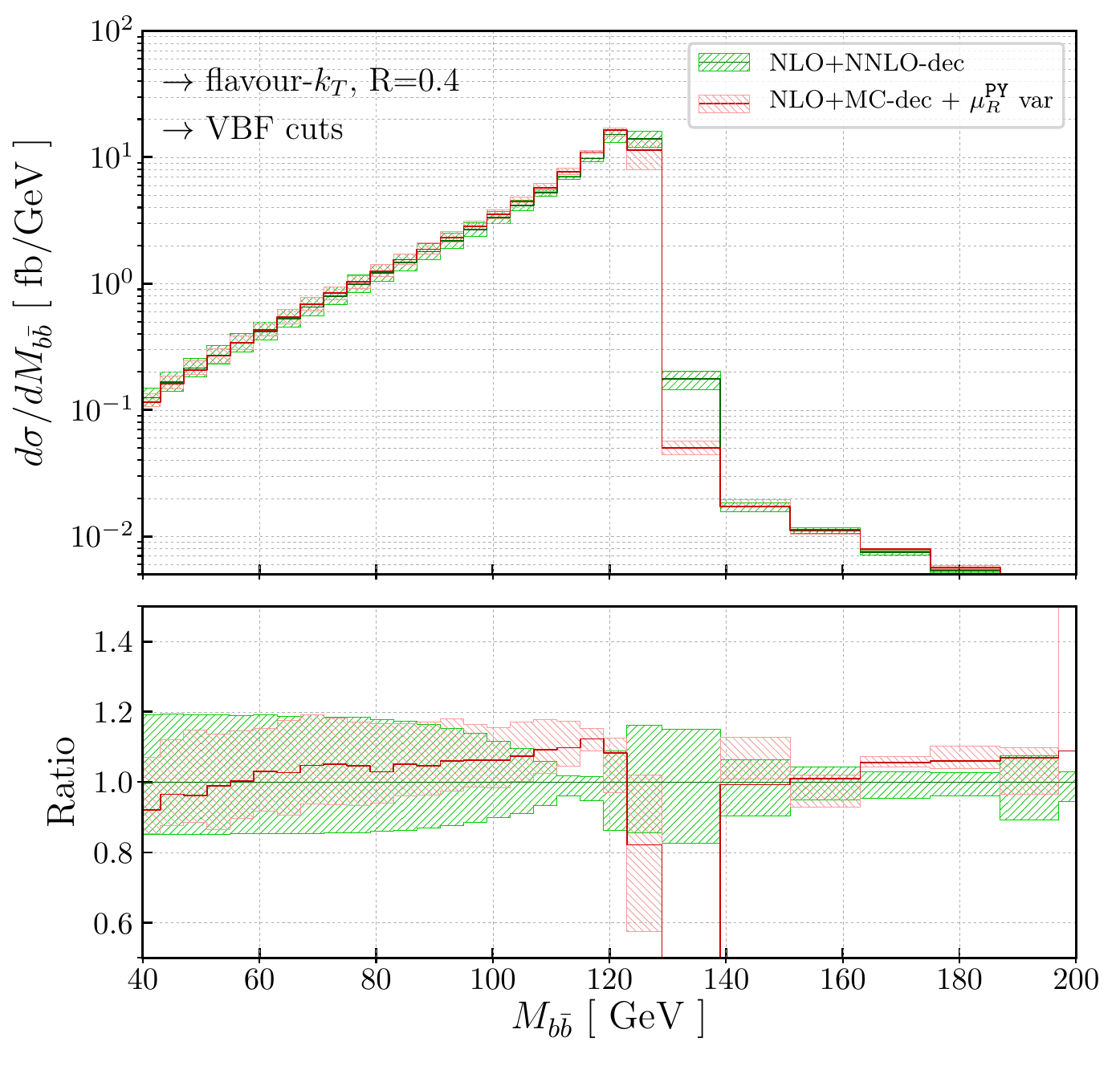}\hfill
\includegraphics[width=0.45\linewidth]{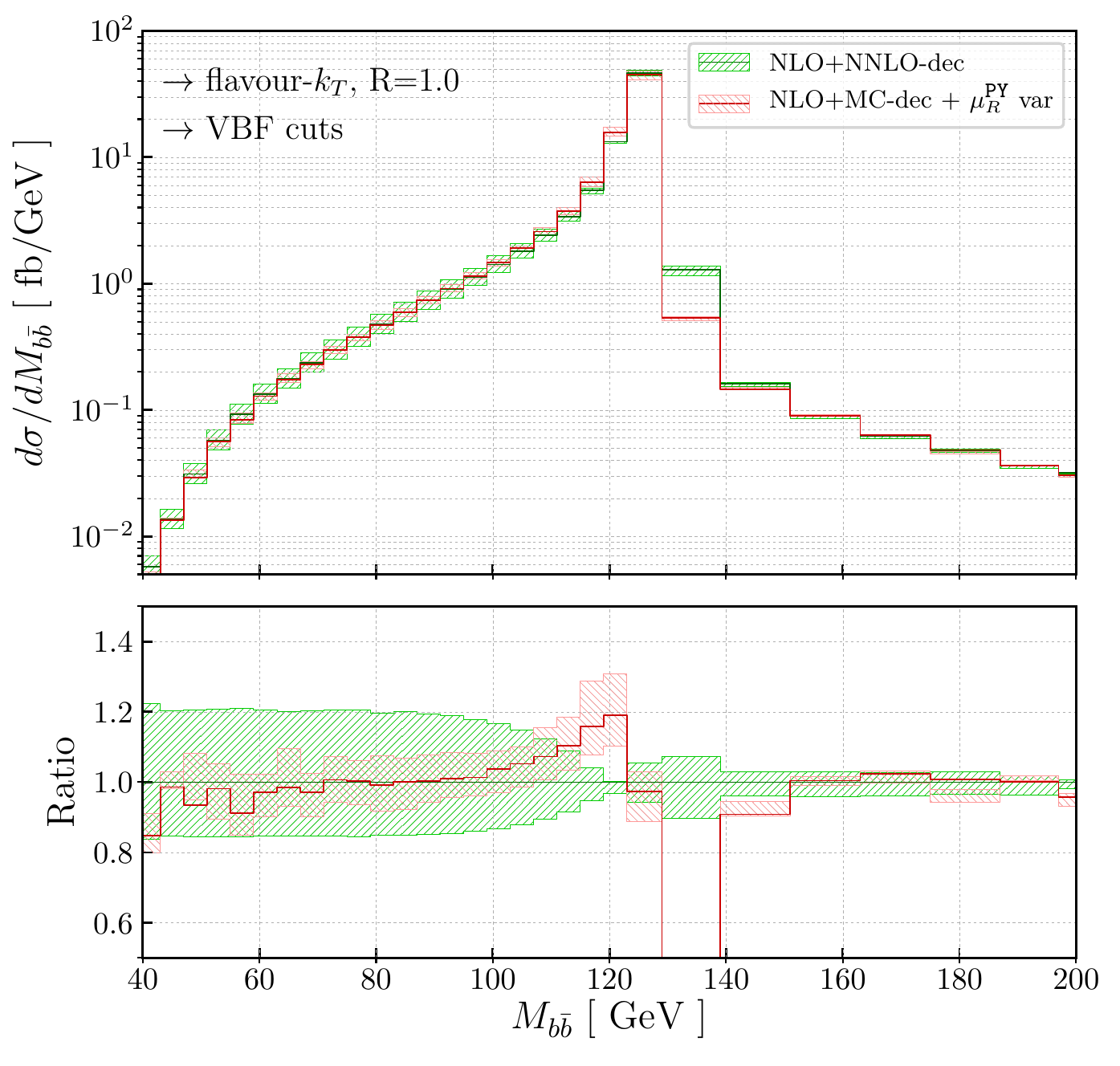}
\caption{Same results as in Fig.~\ref{fig:vbf_mbb}, but here the
  ``NLO+MC-dec'' uncertainty band contains also PS scale variation, as explained in the text.}
\label{fig:vbf_mbb_PSvar}
\end{figure}

\begin{figure}
\centering 
\includegraphics[width=0.45\linewidth]{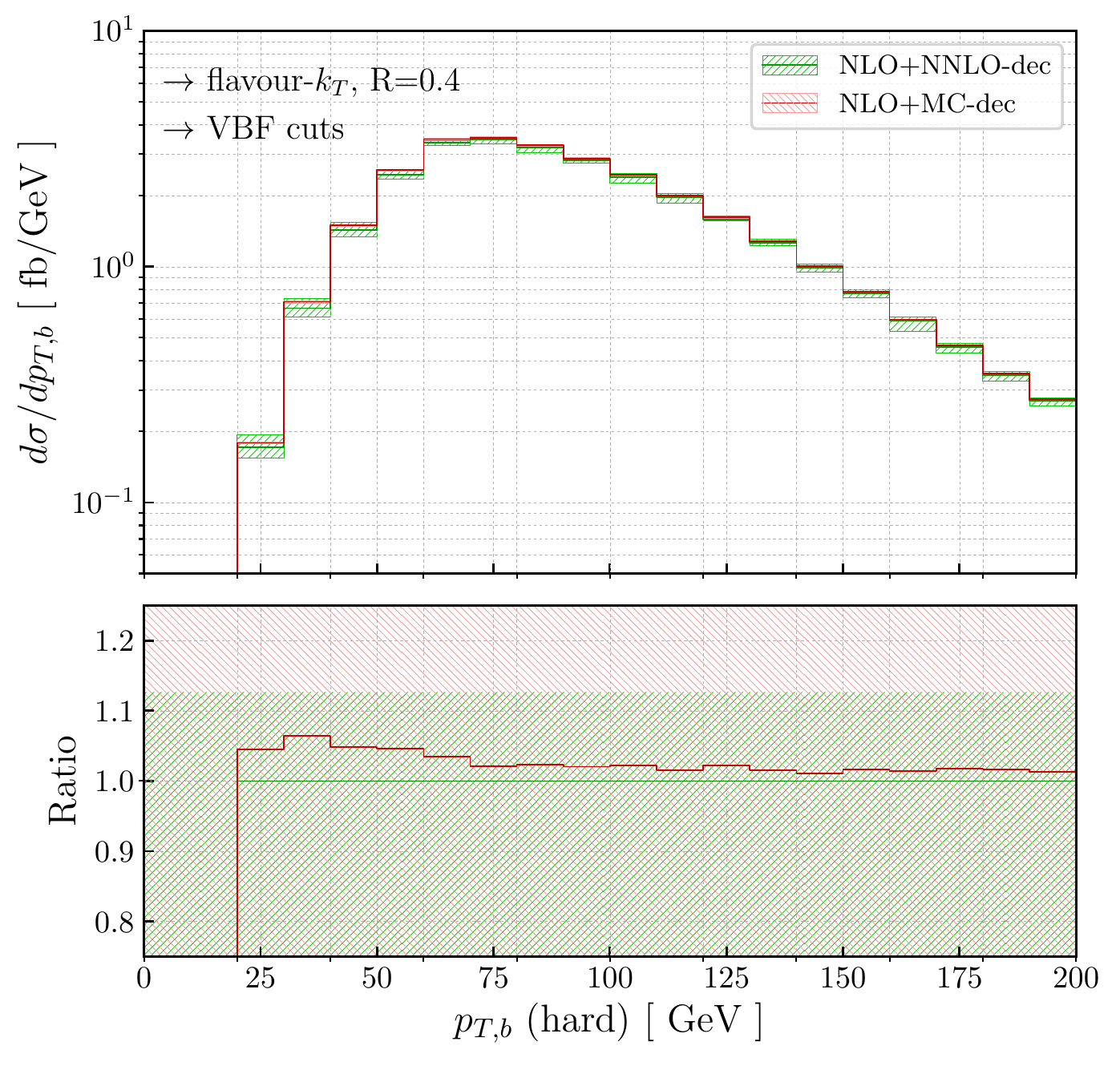}\hfill
\includegraphics[width=0.45\linewidth]{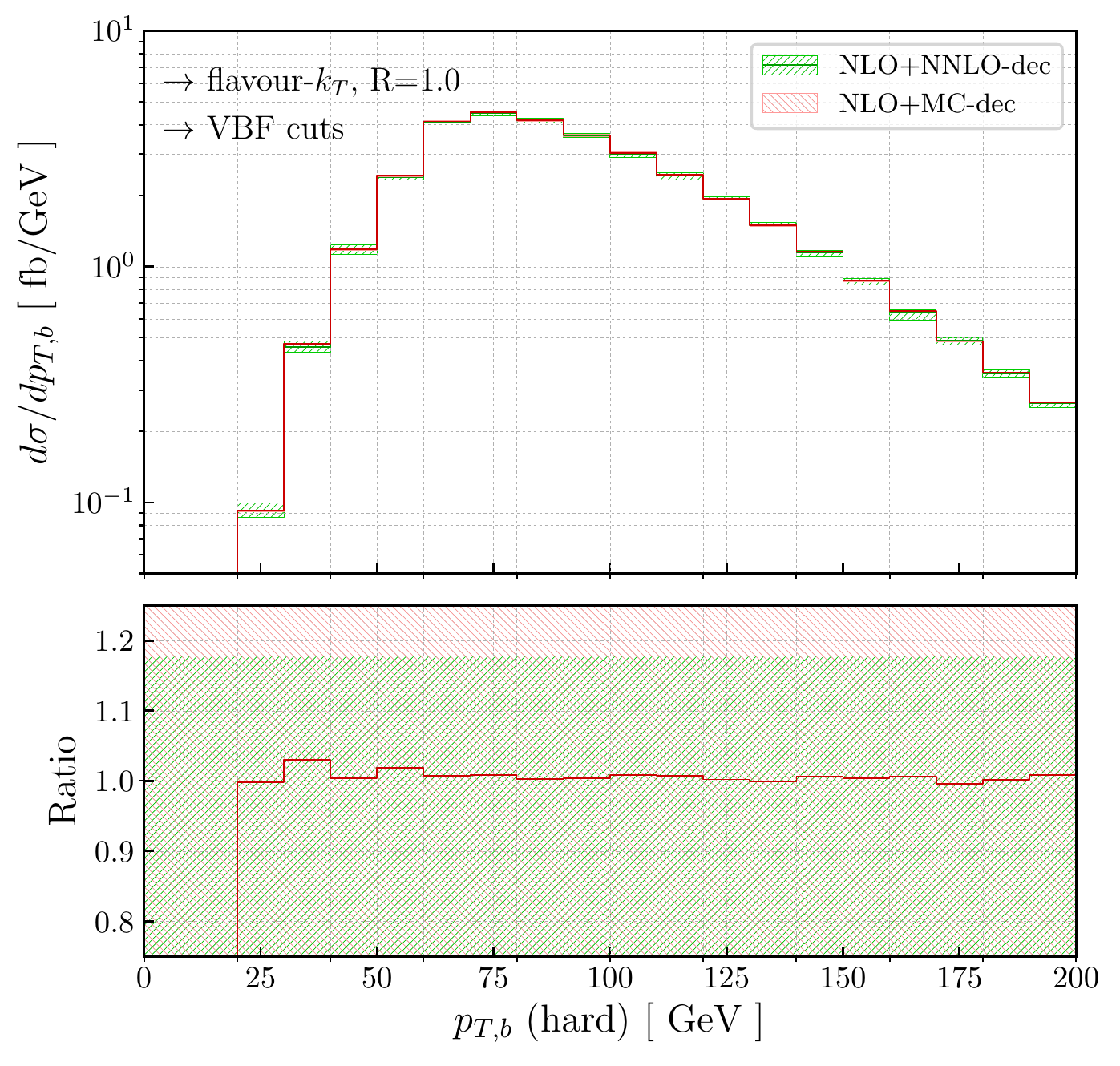}
\caption{Transverse momentum of the hardest $b$-jet in VBF Higgs
  events with Higgs decaying to $b$-quarks. The VBF cuts described in
  the text are applied. The jets are reconstructed with the flavour
  algorithm of Ref.~\cite{Banfi:2006hf} for $R=0.4$ (left) and $R=1.0$
  (right).}
\label{fig:vbf_ptb}
\end{figure}

In Fig.~\ref{fig:vbf_mbb_PSvar} we address this fact more in detail:
we show the ``NLO+MC-dec'' result including scale variation within
\PYTHIA{8}, and we compare it with the uncertainty of the
``NLO+NNLO-dec'' result.
To this end, we use the ``automated parton-shower variations''
facility of \PYTHIA{8}, which was introduced in
Ref.~\cite{Mrenna:2016sih}. In particular, in order to show a fair
comparison with the ``NLO+NNLO-dec'' result, we only show the effect
obtained from changing the renormalisation scale for final-state
radiation (FSR) by a factor of $1/2$ (\texttt{fsr:muRfac=0.5}) and $2$
(\texttt{fsr:muRfac=2.0}). We also correlate this variation with the
one performed in the \POWHEG{} simulation of VBF events.\footnote{We
  do not include the effect of scale variation within \PYTHIA{8} on
  the ``NLO+NNLO-dec'' result. This goes well beyond the scope of this
  work, as it would require to study the interplay between the
  mechanisms that generate the first two radiations off the $b$-quarks
  from the Higgs decay (\hbbgMINLO{}), and the mechanism of PS scale
  variation in \PYTHIA{8}, which, in its current form, is not designed
  to deal with such a complex situation. Having said that, as far as
  the region $\mbb\lesssim\mh$ is concerned, we expect that the
  dominant uncertainty of the ``NLO+NNLO-dec'' will be driven by scale
  variation witin the \hbbgMINLO{} generator.} The corresponding band
is shown, in red, in Fig.~\ref{fig:vbf_mbb_PSvar}, where the
``NLO+NNLO-dec'' results (green) are exactly as in
Fig.~\ref{fig:vbf_mbb}.
We observe that the relative uncertainty of the ``NLO+MC-dec'' result
becomes now more realistic, yielding about $\pm 10\%$ for $\mbb
\lesssim \mh$. The ``NLO+NNLO-dec'' uncertainty reaches instead
$+20\%$ and $-15\%$ for $\mbb \leq 80~\gev$.
Although a comprehensive assessment of PS uncertainties goes beyond
the scope of this paper, we conclude that, when PS scale variation is
included, the uncertainties of the ``NLO+NNLO-dec'' and the
``NLO+MC-dec'' are of the same order. For small $\mbb$ values,
the ``NLO+NNLO-dec'' displays a wider uncertainty band, which is
especially visible when the jet radius is large.

Next, in Fig.~\ref{fig:vbf_ptb}, we present the distributions of the
transverse momentum of the hardest $b$-jet.
Both results are in very good agreement. Nevertheless, we observe a
slight distortion of the ``NLO+NNLO-dec'' result with respect to the
``NLO+MC-dec'' one for small values of the $p_{t, b}$(hard) in case of
$R=0.4$ jet radius (Fig.~\ref{fig:vbf_ptb}, left plot).
Once again, this can be attributed to out-of-cone radiation which is
more pronounced in the ``NLO-NNLO-dec'' result. Such an effect causes
events with moderate initial value of the $p_{t,b}$ to fail to
fulfill fiducial requirements after the QCD radiation due to the NNLO
treatment and PS evolution.
We also notice that the scale uncertainty of the ``NLO+NNLO-dec'' band
widens in the region of small transverse momenta consistently with the
fact that in this region perturbative uncertainties are bigger than
for the hard part of the spectrum.

\section{Conclusions}
\label{sec:conclu}

In this work, we presented a simple method to describe the $\hbb{}$
decay with state-of-the-art NNLO accuracy consistently matched to a
parton shower.
We also provide a simple tool that allows to combine decay
events with any available (N)NLOPS description of a Higgs production
process.

The method exploits the factorization between production and decay of
the Higgs boson, and relies on generating Higgs production events at
(N)NLO accuracy, where the Higgs boson is undecayed, and combining
them with Higgs decay events that describe the decay of the Higgs boson 
to $b$-quarks at NNLO accuracy.
These combined LH events can then be showered using any transverse momentum
ordered parton shower. Since radiation from production and decay factorize,
the parton shower requires two separate upper bounds for the
radiation, one for production and one for the decay. The former one is
stored in the event file, while the latter can be easily computed from
the decay kinematics.

As a proof of principle, we studied the effect of including NNLOPS
corrections to the decay in the case of HZ Higgs production
and VBF
Higgs production processes.
We find that parton shower effects are moderate once NNLO corrections
to the decay are included. Nevertheless, including higher-order
corrections is important since it enables one to use information
encoded in the NNLO accurate matrix elements during event generation.
We have released the source code of the program, as an {\texttt{h\_bbg}}
user process in the \POWHEGBOXVTWO{}, that enables a user to generate
\hbbgMINLO{} events and further reweight them such that the NNLO
\hbb{} decay width is reproduced.
Although the generation process is straightforward and extremely fast,
a ready-to-use sample of event files is also available upon request.
We also include a tool to merge decay events with a Higgs production
mode and a parton shower driver that allows for a consistent interface
with \PYTHIA{}.
At the moment, the code can be interfaced to any single-Higgs
production mode but extension to double-Higgs production, where one or
both Higgs bosons decay to $b$-quarks, is possible.

Currently, the NNLO corrections to the decay are treated in the
massless approximation. Nevertheless, $b$-quark mass effects to the
\hbb{} decay can also be included in a similar manner.
This step is important for obtaining a consistent treatment of the
$b$-quark mass in the \hbbgMINLO{} generator as well as in the parton
shower evolution. We leave that for future work.

\section*{Acknowledgments}
We wish to thank Pier Francesco Monni for numerous useful discussions.
We are grateful to and G\'abor Somogyi and Claude Duhr for providing
the analytic expressions of the $\hbb{g}$ amplitudes.
We also thank Paolo Nason for discussions about \POWHEG{} processes
with no initial state partons.
This work was supported in part by ERC Consolidator Grant HICCUP (No.\
614577). The work of ER was partially supported by a Marie
Sk\l{}odowska-Curie Individual Fellowship of the European Commission's
Horizon 2020 Programme under contract number 659147
PrecisionTools4LHC.
WB and ER thank the Max Planck Institute for Physics for hospitality
while part of this work was carried out.
ER thanks also the Rudolf Peierls Centre for Theoretical Physics at
the University of Oxford for hospitality.

\appendix
\section{\MINLO{} Sudakov}
\label{app:resformulae}
In this appendix we discuss how to construct a \MINLO{} Sudakov form
factor, specifically how to obtain the process dependent $B_2$
coefficient that also accommodates effects related to the
multiple emissions.

The Sudakov radiator can be written as
$\Delta(\yres) = \exp\{\, R(\yres) \}$, where the function $R(\yres)$
can be parametrised as follows
\begin{align}
  R(\yres) = - L g_1(x) - g_2(x) - \frac{\as}{\pi} g_3(x),
\end{align}
where $L = \ln(1/\yres)$ is the logarithm that has been resummed, $x =
\as \beta_0 L$, and the strong coupling $\as$ is evaluated at the hard scale
$M_H$.
The functions $g_1$, $g_2$ and $g_3$ can be obtained from the App.~B of
Ref.~\cite{Banfi:2014sua}, by setting $a=2$ and taking the limit
$b_{\ell}\to 0$. They read
\begin{align}
  g_1(x)
  =&
     \frac{A_1}{(2\pi\beta_0)}\,\frac{x+\log(x)}{2x}\,,
     \nonumber\\
  g_2(x)
  =&
     -\frac{A_2}{(2\pi\beta_0)^2} \lp \frac{x}{\omx} + \log(\omx) \rp
     + \frac{B_1}{(2\pi\beta_0)} \log(\omx)
     \nonumber\\
     & + \frac{A_1}{(2\pi\beta_0)}
     \frac{\beta_1}{\beta_0{\,^2}}\lp
     \frac{x + \log(\omx)}{\omx} + \frac{1}{2}\log^2(\omx)\rp\,,
  \nonumber\\
  g_3(x)
  =&
     \frac{1}{2(2\pi\beta_0)^3}\bigg[
     -\frac{B_2}{(2\pi\beta_0)} \frac{x}{2(\omx)}
     +\frac{A_2}{(2\pi\beta_0)^2} \frac{\pi\beta_1}{\beta_0}\lp\lp-1+\tfrac{3}{2}x\rp+(-1+2x)\frac{\log(\omx)}{x}\rp
     \nonumber \\
   & + \frac{B_1}{(2\pi\beta_0)} \frac{\beta_1}{\beta_0} \frac{x+\log(\omx)}{\omx}
     + \frac{A_1}{(2\pi\beta_0)} \frac{\pi\beta_2}{\beta_0^{\,2}} \lp \frac{(1-\tfrac{3}{2}x)x}{(\omx)^2} + \log(\omx) \rp
     \nonumber \\
   & + \frac{A_1}{(2\pi\beta_0)} \frac{\pi\beta_1^{\,2}}{\beta_0^{\,3}} \lp \frac{x^2+\log^2(\omx)}{2(\omx)^2} + \frac{x(\omx-\log(\omx))\log(\omx)}{(\omx)^2}\rp 
     \bigg].
\end{align}
In the expressions above, the constant part of the radiator
(i.e. $R(\yres\to0)$) has been removed, as these corrections are
already reconstructed by the $H\to 4~$partons matrix elements of the real
corrections to the \hbb{g} process. Moreover, we also neglect terms
proportional to the $A_3$ coefficient as they give rise to terms 
$\mathcal{O}(\as^2)$ terms after integration over radiation which are beyond our accuracy.

The values of the $\beta$-function coefficients read
\begin{align}
  \beta_0 =& \frac{11 C_A - 2 N_f}{12\pi},
  \nonumber\\
  \beta_1 =& \frac{17C_A^2 - 5C_A N_f - 3C_F N_f}{24\pi^2}
  \nonumber\\
  \beta_2 =& \frac{
  2857 C_A^3
  +(54 C_F^2-615 C_F C_A-1415 C_A^2) N_f
  +(66 C_F+79 C_A) N_f^2
  }{3456 \pi^3}\,.
\end{align}

The relevant anomalous dimensions read
\begin{align}
  A_1 &= 2 C_F, &
  B_1 &= -3 C_F, &
  A_2 &= C_F K_{\scriptscriptstyle\rm{CMW}},
\end{align}
where
\begin{align}
  K_{\scriptscriptstyle\rm{CMW}}
  =&
     C_A \lp \frac{67}{9} - \frac{\pi^2}{3} \rp
     - \frac{10}{9}N_f,
\end{align}
with $N_f$ being a number of light quark flavours.

The $B_2$ anomalous dimension requires a slightly more detailed
treatment. Our starting point is the Drell-Yan
value~\cite{Davies:1984hs,deFlorian:2001zd}
\begin{align}
  B_2
  =&
     -2\bigg(
     \lp -3\zeta_2 + \frac{3}{8} + 6\zeta_3 \rp C_F^2
     + \lp \frac{11}{3}\zeta_2 + \frac{17}{24} - 3\zeta_3 \rp C_F C_A
     \nonumber \\
   &+ \lp -\frac{1}{12} - \frac{2}{3}\zeta_2 \rp C_F N_f
     \bigg)
     + (2\pi\beta_0) C_F \zeta_2.
\end{align}
First, we need to replace the $\mathcal{O}(\as)$ $Z\to qq$ virtual
corrections with the virtual corrections of the \hbb{}
process~\cite{DelDuca:2015zqa}. This is achieved with the replacement
\begin{align}
  B_2 \longrightarrow
  B_2 + 2\pi\beta_0 \lp \pi^2 - 2 + \zeta_2 \rp.
\end{align}
Furthermore, we need to include the corrections due to resolved real
radiation in various kinematical configurations~\cite{Banfi:2016zlc}, i.e
\begin{align}
  \delta\mathcal{F} =
  \delta\mathcal{F}_{\rm sc.}
  +\delta\mathcal{F}_{\rm clust.}
  +\delta\mathcal{F}_{\rm correl.}
  +\delta\mathcal{F}_{\rm hc.}
  +\delta\mathcal{F}_{\rm rec.}
  +\delta\mathcal{F}_{\rm wa.},
\label{eq:F}
\end{align}
where the soft-collinear term ($\delta\mathcal{F}_{\rm sc.}$) comes
from the running coupling effects of the matrix element for soft
emission as well as keeping the correct phase space boundary for a
soft emission; the clustering and correlated corrections
($\delta\mathcal{F}_{\rm clust.}$ and
$\delta\mathcal{F}_{\rm correl.}$) are a result of the two soft
emissions being close in rapidity, affecting the clustering sequence
and arising from non-abelian correlations; the hard-collinear and
recoil terms ($\delta\mathcal{F}_{\rm hc.}$ and
$\delta\mathcal{F}_{\rm rec.}$) are responsible for treating the
configurations with collinear but hard emission by taking the correct
approximation of the matrix element (hard-collinear) as well as
treating the recoil effects in the observable; finally the wide-angle
correction ($\delta\mathcal{F}_{\rm wa.}$) deals with emission which
is soft but emitted at wide angle with respect to the emitter.  In
case of the C/A clustering algorithm the soft-collinear and
hard-collinear corrections disappear,
\begin{align}
  \delta\mathcal{F}_{\rm sc.} =&\, 0, & \delta\mathcal{F}_{\rm hc.} =&\, 0.
\end{align}
Furthermore, the recoil and wide-angle corrections are proportional to
$\as(q_t)$ (see Eqs.~(48) and (53) of Ref.~\cite{Banfi:2016zlc}) and
hence are already taken into account in the matrix elements for the
real corrections, that directly enter the $\bar{B}$ function. Note
that the \MINLO{} choice for the scale of the strong coupling is
essential in here.
The remaining two terms, $\delta\mathcal{F}_{\rm clust.}$ and
$\delta\mathcal{F}_{\rm correl.}$, scale as $\as^2(q_t) L$ and need
to be taken into account at the level of Sudakov radiator. We use the
results reported in Eqs.~(34) and (39) of
Ref.~\cite{Banfi:2016zlc}. We have
\begin{align}
  \delta\mathcal{F}_{\rm clust.}
  =&
     \sum_{\ell=1,2} \as^2(q_t)L \lp \frac{C_F}{\pi}\rp^2 \mathcal{I}_{\rm clust.}
     + \mathcal{O}(\as^3),
  \\
  \delta\mathcal{F}_{\rm correl.}
  =&
     \sum_{\ell=1,2} \as^2(q_t)L \lp\frac{C_F}{2\pi^2}\rp \mathcal{I}_{\rm correl.}
     + \mathcal{O}(\as^3),
\end{align}
where the integrals $\mathcal{I}_{\rm clust.}$ and
$\mathcal{I}_{\rm correl.}$ are
\begin{align}
  \mathcal{I}_{\rm clust.} &\approx -0.493943, \nonumber\\
  \mathcal{I}_{\rm correl.} &\approx 2.1011 C_A + 0.01496 N_f.
\end{align}
Since ${\cal F}$ in Eq.~\eqref{eq:F} simply multiplies the Sudakov
form factor, these corrections can be included as a further shift in
$B_2$
\begin{align}
  \label{eq:fshift}
  B_2 \longrightarrow
  B_2 - (2\pi)^2 ( \delta F_{\rm clust.} + \delta F_{\rm correl.} ),
\end{align}
with $\delta F_{i} = \delta\mathcal{F}_{i} / (\as^2(q_t)L)$.

We are now ready to assemble the Sudakov form factor used in
Eq.~\eqref{eq:bbar}.
Furthermore, in order to probe the uncertainty related to missing
higher-order terms, the renormalisation scale dependence in the
Sudakov radiator $R(\yres)$, that is $\as(\mh) \to \as(\mur)$ change,
can be implemented along the lines of Eqs.~(B.7)-(B.9) of
Ref.~\cite{Banfi:2016zlc}. Similarly, we can introduce the resummation
scale dependence that allows to estimate the uncertainty related to missing 
higher-order logarithmic terms, i.e.
\begin{align}
  L \longrightarrow L = \ln( \kappa_{\rm res}^{\,2} / \yres),
\end{align}
with $\kappa_{\rm res} = Q_{\rm res} / \mh$. This shift is compensated by the following 
changes in the $g_i(x)$ functions, which guarantee that the logarithmic accuracy is preserved: 
\begin{align}
  g_1(x) \longrightarrow&\;
  g_1(x),
  \nonumber \\
  g_2(x) \longrightarrow&\;
  g_2(x) + (A_1\pi\beta_0) \frac{x}{\omx} \log\lp\kappa_{\rm res}^{\,2}\rp,
  \\
  g_3(x) \longrightarrow&\;
  g_3(x)
  - A_1\frac{(\pi\beta_0)^3}{2(\omx)^2} \log^2\lp\kappa_{\rm res}^{\,2}\rp
  \nonumber \\
  &+ \bigg[
    A_2(\pi\beta_0)^2\frac{x}{(\omx)^2}
    + B_1 (\pi\beta_0)^3 \frac{2}{\omx}
    - 2A_1(\pi^3\beta_0\beta_1) \frac{x\log(\omx)}{(\omx)^2}
    \bigg] \log\lp\kappa_{\rm res}^{\,2}\rp.
  \nonumber
\end{align}
We set the value of the central resummation scale to be $Q_{\rm res} =
\mh$.
For the results presented in this work the resummation scale was not
varied, i.e. $\kappa_{\rm res} = 1$.

\section{Analytic proof of NLO accuracy of \hbbgMINLO{} for $\hbb{}$}
\label{app:analyticproof}

We start from the resummation formula for $y_3$ in $\hbb{}$ decay that
is matched to the NLO $\hbb{}$ decay width $\Gamma_{\bb}^{(1)}$, i.e.
\begin{align}
  \Sigma(y_3)
  ={}&
  \frac{1}{\Gamma_{\bb}^{(0)}}
  \int_0^{y_3} dy'_3
  \, \frac{d\Gamma_{\bb}}{d y'_3}\,,
\end{align}
and it can be parametrised as
\begin{equation}
  \Sigma(y_3) = C_{\bb}(\as) e^{R(\as,y_3)} + \Sigma_{\rm fin}(\as,y_3)
        \,,
\label{eq:Sigma}
\end{equation}
where 
\begin{align}
  C_{\bb}(\as)
  ={}&
  1 + \as  C_{\bb}^{(1)} + \mathcal{O}(\as^2)
  \,,
  \\
  R(\as,y_3)
  ={}&
  \as R^{(1)}(y_3)
  + \as^2 R^{(2)}(y_3)
  + \mathcal{O}(\as^3)
  \,,
  \\
  \Sigma_{\rm fin}(\as,y_3)
  ={}&
  \as \Sigma_{\rm fin}^{(1)}(y_3) + \mathcal{O}(\as^2)
  \,.
\end{align}
As discussed in App.~\ref{app:resformulae}, the radiator functions
include all logarithms related to the exponentiation of the one-gluon
result, but also all NNLL real-radiation multiple emission effects
which are incorporated through a modification of the $B_2$
coefficient. $\Sigma^{(1)}_{\rm fin}(y_3)$ contains the ${\cal
  O}(\alpha_s)$ part of the cross section which is finite and vanishes
for $y_3\to0$. Explicitly one has
\begin{equation}
  \Sigma^{(1)}_{\rm fin}(y_3)  = \Sigma^{(1)}(y_3)- R^{(1)}(y_3) - C_{\bb}^{(1)} 
  \,. 
\end{equation}

By definition, we have that $\Sigma(y_{\rm max}) =
(\Gamma_{\bb}^{(0)}+\as \Gamma_{\bb}^{(1)})/\Gamma_{\bb}^{(0)} + {\cal O}(\as^2)$. This
means that if one integrates the distribution up to the maximum
kinematically allowed value of $y_3$, one obtains, by construction,
the NLO K-factor of the $\hbb{}$ decay width, regardless of the form
of the Sudakov form factor.

Taking the derivative of Eq.~\eqref{eq:Sigma} one obtains
\begin{equation}
  \frac{d\Sigma(y_3)}{d L}
  = C_{\bb}(\as)
e^{R(y_3)} \frac{dR(\as,y_3)}{d L} +
\frac{d\Sigma_{\rm fin}(\as,y_3)}{d L}
\,,
\label{eq:dSigma}
\end{equation}
where $L = \log(1/y_3)$.
We then define
\begin{equation}
  D(\as,y_3) = C_{\bb}(\as) \frac{dR(\as,y_3)}{d L}
+ e^{-R(\as,y_3)}\frac{d\Sigma_{\rm fin}(\as,y_3)}{d L}
  \,,
\label{eq:D}
\end{equation}
and expand $D(y_3)$ in powers of $\as= \as(y_3 Q^2)$ arriving at
\begin{equation}
D(\as,y_3) = \as D^{(1)}(y_3)+\as^2 D^{(2)}(y_3)+\mathcal{O}(\as^3) \,.
\label{eq:Dexp}
\end{equation}
It is clear that if one integrates Eq.~\eqref{eq:dSigma} over the
whole range of $y_3$, one obtains back the NLO K-factor of the $\hbb{}$ decay
width.

In order to see which terms can be dropped without spoiling the
NLO accuracy of the result, we remind the reader that the
power-counting is fixed by the integral
\begin{equation}
\int dL \, \as^n L^m e^{R(\as,y_3)} = {\cal O}\left(\as(Q^2)^{n-\frac{m+1}{2}}\right)\,.  
\label{eq:counting}
\end{equation}
It is then easy to see that, provided the expansion is done using
$\as=\as(y_3 Q^2)$, the largest term beyond ${\cal O}(\as^2)$ comes
from the $A_3$ coefficient in the Sudakov form factor.
After taking the derivative, these terms contribute as $\as^3
L$. Hence, upon integration, according to Eq.~\eqref{eq:counting} they
give rise to an NNLO correction. This means that all ${\cal O}(\as^3)$
terms in $D(\as,y_3)$ can be dropped without spoiling the NLO accuracy of
the result after integration. From Eq.~\eqref{eq:D} it is clear that
$D(\as,y_3)$ also contains a term $\as^2 B_2 $, which upon integration
gives a correction ${\cal O}(\as^{3/2})$ and therefore has to be
included.
%
On the other hand Eq.~\eqref{eq:dSigma} can be expanded in powers of the coupling constant as 
\begin{equation}
  \frac{d\Sigma(y_3)}{d L}
  = \as \frac{d\Sigma^{(1)}(y_3)}{d L} + \as^2 \frac{d\Sigma^{(2)}(y_3)}{d L} + {\cal O}(\as^3) .
\end{equation} 
By explicitly taking the derivative of Eq.~\eqref{eq:Sigma}, the above terms
can be writen as
\begin{equation}
\frac{d\Sigma^{(1)}(y_3)}{d L}   = \frac{dR^{(1)}(y_3)}{d L} = D^{(1)}(y_3)
\end{equation}
and
\begin{equation}
\frac{d\Sigma^{(2)}(y_3)}{d L}  = R^{(1)}(y_3) D^{(1)}(y_3) +D^{(2)}(y_3)\,. 
\end{equation}

Accordingly, dropping terms that are order $\as^3$ and higher, one can
rewrite Eq.~\eqref{eq:dSigma} as
\begin{equation}
  \frac{d\Sigma(y_3)}{d L}
  ={}
  e^{R(\as,y_3)} \left( 
  \as\big(1-\as R^{(1)}(y_3)\big) \frac{d\Sigma^{(1)}(y_3)}{d L} 
  +\as^2 \frac{d\Sigma^{(2)}(y_3)}{d L} 
  \right)\,.
  \label{eq:dSigma2}
\end{equation}
We now clearly see that Eq.~\eqref{eq:dSigma2} agrees with the
standard \MINLO{} formula, which then also must have NLO accuracy
provided the Sudakov form factor includes the $B_2$ resummation coefficient.
We stress that it is crucial that the expansion in Eq.~\eqref{eq:Dexp}
is performed around $\as=\as(y_3 Q^2)$. Would one expand around
$\as(Q^2)$ then ${\cal O}(\as^3 L^2)$ terms would be present in
Eq.~\eqref{eq:Dexp} which would contribute as ${\cal O}(\as^{3/2})$.
Accordingly, this sets the scale of the strong coupling constant in
the \MINLO{} framework to $\mu^2 = y_3 Q^2$.

\section{$b$-quark mass effects in the parton shower}
\label{app:mb0}
\begin{figure}[h]
\centering 
\includegraphics[width=0.45\linewidth]{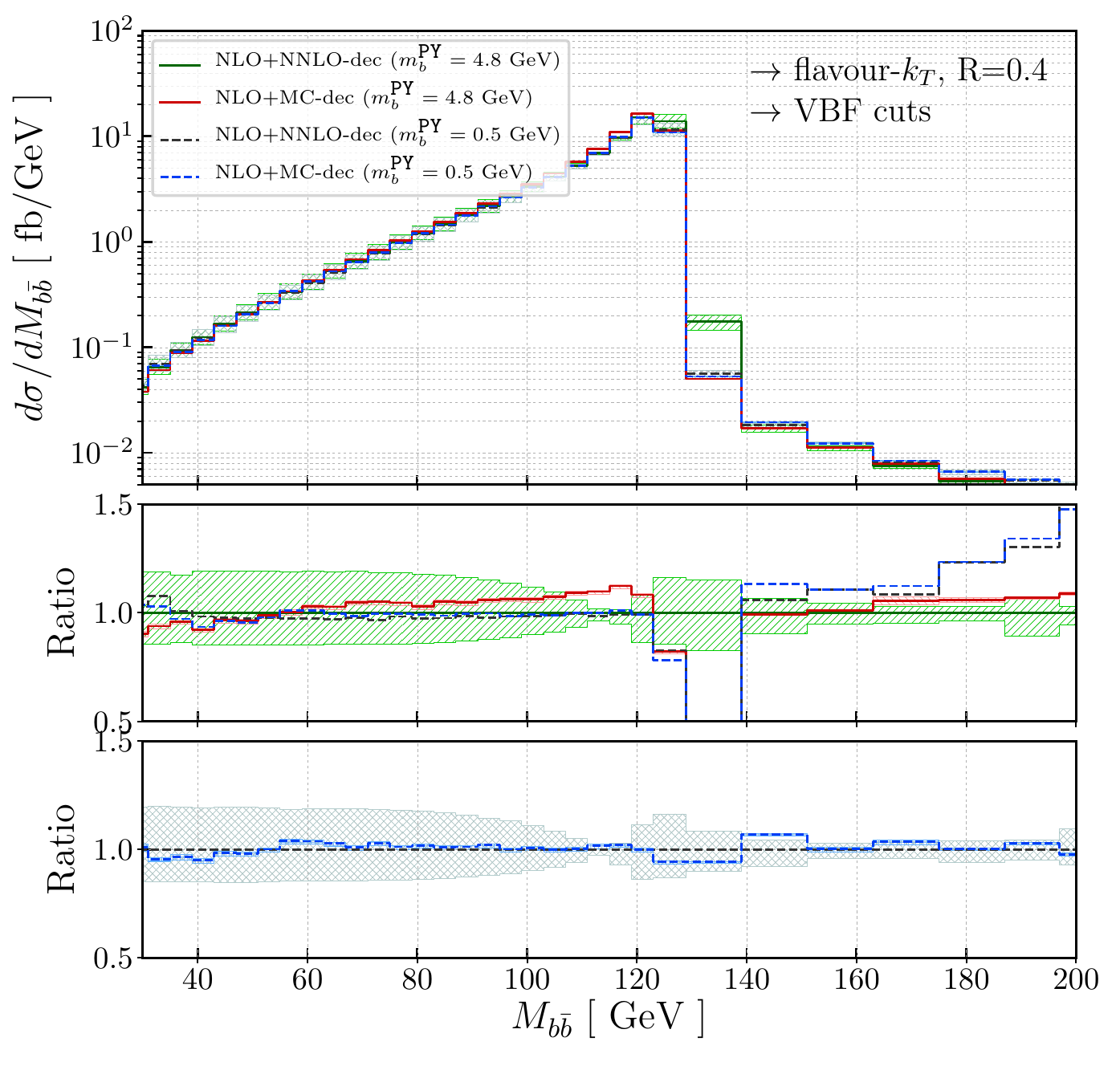}\hfill
\includegraphics[width=0.45\linewidth]{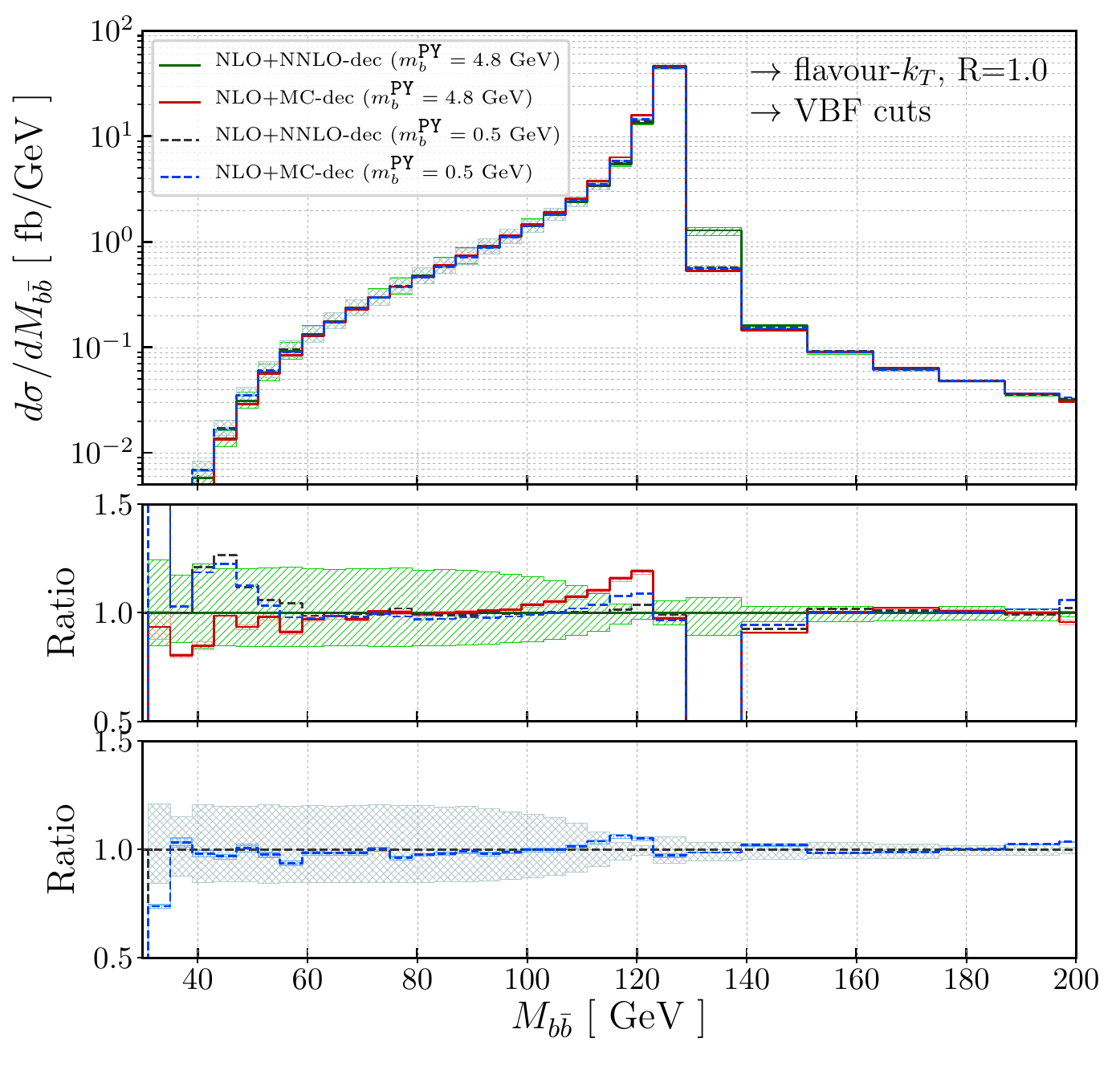}\hfill
\caption{Invariant mass of the two $b$-jets in VBF Higgs events with
  Higgs decaying to $b$-quarks. The VBF cuts described in
  Sec~\ref{sec:cutsvbf} are applied. The jets are reconstructed with
  the flavour algorithm of Ref.~\cite{Banfi:2006hf} for $R=0.4$ (left)
  and $R=1.0$ (right). The plots show the dependence of the results
  when approaching the small $b$-quark mass in \PYTHIA{8}, see text
  for details.}
\label{fig:vbf_mbb_mbPY_2}
\end{figure}

As anticipated in Sec.~\ref{sec:vbf}, for the VBF case we have
investigated the differences due to the treatment of the $b$-quark
mass in more detail.
For this purpose, in Fig.~\ref{fig:vbf_mbb_mbPY_2}, we show the
distribution of the invariant mass $\mbb$ obtained with the default
\PYTHIA{8} setting ($\mbPY = 4.8~\gev$, solid lines) as well as with
approximately massless $b$ quarks in \PYTHIA{8} ($\mbPY = 0.5~\gev$,
dashed lines).

For both setups, we show the ``NLO+NNLO-dec'' (green for $\mbPY =
4.8~\gev$, black dashes for $\mbPY = 0.5~\gev$) and the ``NLO+MC-dec''
results (red for $\mbPY = 4.8~\gev$, blue dashes for $\mbPY = 0.5~\gev$).
The main panel of Fig.~\ref{fig:vbf_mbb_mbPY_2} shows the absolute
value of the cross section. The middle panel shows the ratio of each
prediction with respect to the ``NLO+NNLO-dec'' result obtained with
$\mbPY = 4.8~\gev$. For simplicity, we don't show in this panel the
uncertainty band of the results obtained with $\mbPY = 0.5~\gev$. The
bottom panel shows instead the ratio between the ``NLO+MC-dec''
central prediction (and uncertainty band) for $\mbPY = 0.5~\gev$,
normalized with respect to the ``NLO+NNLO-dec'', obtained as well with
$\mbPY = 0.5~\gev$.

We first focus on the region $\mbb \lesssim \mh$.
For $R=0.4$ (left plot) we see that, once $\mbPY$ is set to a small
value, the ``NLO+MC-dec'' result (blue) becomes almost identical to
the ``NLO+NNLO-dec'' results, both with small $\mbPY$ mass (black) and
with the default $\mbPY$ setting (green). Therefore the differences
between the ``NLO+MC-dec'' prediction (red) and the ``NLO+NNLO-dec''
one (green) in this region can be attributed to the treatment of
$b$-mass effects, as already stated in Sec.~\ref{sec:vbf}. The same
conclusion can be drawn for $R=1.0$ (right plot), although for
$\mbb\simeq \mh$ very minor differences between ``NLO+MC-dec'' (blue)
and the ``NLO+NNLO-dec'' (black) predictions remain.
We also observe that, for $R=1.0$, the region $\mbb <  60$ GeV
displays more pronounced differences. This region is very suppressed
and the distribution is very steeply falling. It's not surprising that
a prediction where the $b$-mass is kept equal to its physical value falls
off faster.
We observe a similar effect for $R=0.4$ in the region $\mbb >
180~\gev$ where the cross section falls off very fast, signalling that
one is approaching the edge of the available phase space, and
therefore differences between simulations are to be expected.

The most striking difference for $\mbb > \mh$ is just to the right of
the peak. Here we observe that the ``NLO+NNLO-dec'' prediction with
massive $b$-quarks in \PYTHIA{8} (green) differs from all the other
results.
Most likely this difference can be attributed to how one calculates
the hardness and the PS veto condition. The \hbbgMINLO{} generator
computes the radiation upper bound \texttt{scalup} using the $m_b=0$
approximation in the matrix elements and in the \MINLO{}
Sudakov. Nevertheless, the parton shower evolution is carried out with
$b$-quarks which were assigned a mass $\mbPY \neq 0$. As the radiation
pattern of massive and massless quarks is slightly different, one can
expect a small mismatch due to this. Once again, this confirms that
the region close to the peak is very sensitive to how $b$-mass effects
are taken into account.
We conclude that, in the future, it would be desireable to include
$b$-mass effects in all parts of the NNLOPS simulation, in particular
in the \hbbgMINLO{} generator. Such a study goes well beyond the scope
of this work, and it should be addressed in the future.

\bibliography{hbbg}{}
\bibliographystyle{JHEP}

\end{document}